\pdfoutput=1 
\documentclass[pre,footinbib,twocolumn,final,superscriptaddress]{revtex4}

\usepackage{amssymb,amsbsy,amsfonts,mathrsfs,amsmath}

\usepackage{bbm}

\usepackage[T1]{fontenc} 
\usepackage[english,francais]{babel}

\usepackage{graphicx,epsfig,color}

\usepackage{url,hyperref}  

\usepackage{enumitem}

\definecolor{M_Beige}         {rgb}{0.96 , 0.96 , 0.86}

\definecolor{M_Brown}         {rgb}{0.65 , 0.16 , 0.16}

\definecolor{M_Gold}          {rgb}{1.00 , 0.84 , 0.00}

\definecolor{M_LemonChiffon}  {rgb}{1.00 , 0.98 , 0.80}

\definecolor{M_Orange}        {rgb}{1.00 , 0.60 , 0.00}

\definecolor{M_Pink}          {rgb}{0.80 , 0.55 , 0.60}

\definecolor{M_Violet}          {rgb}{0.83 , 0.21 , 0.93}

\definecolor{M_Green}          {rgb}{0.2 , 0.6 , 0.2}

\definecolor{M_Gray}          {rgb}{0.5 , 0.5 , 0.5}

\definecolor{M_BluPal}          {rgb}{0.7 , 0.7 , 0.9}

\def\Xint#1{\mathchoice
{\XXint\displaystyle\textstyle{#1}}%
{\XXint\textstyle\scriptstyle{#1}}%
{\XXint\scriptstyle\scriptscriptstyle{#1}}%
{\XXint\scriptscriptstyle\scriptscriptstyle{#1}}%
\!\int}
\def\XXint#1#2#3{{\setbox0=\hbox{$#1{#2#3}{\int}$}
\vcenter{\hbox{$#2#3$}}\kern-.5\wd0}}

\def\dashint{\Xint-}

\renewcommand{\leq}{\leqslant}
\renewcommand{\geq}{\geqslant}

\newcommand{\EXP}[1]{\mathrm{e}^{#1}}         

\def\eqdef{\stackrel{\mbox{\tiny def}}{=}}     
\newcommand{\ket}[1]{|\kern.3ex#1\kern.3ex\rangle}
\newcommand{\bra}[1]{\langle\kern.3ex #1 \kern.3ex|}
\newcommand{\mean}[1]{\left\langle #1 \right\rangle} 
\newcommand{\smean}[1]{\langle #1 \rangle} 

\newcommand{\re}{\mathop{\mathrm{Re}}\nolimits}      
\newcommand{\im}{\mathop{\mathrm{Im}}\nolimits}      
\newcommand{\cotg}{\mathop{\mathrm{cotg}}\nolimits}  

\def\I{{\rm i}}                  
\def\D{{\rm d}}                  

\newcommand{\deriv}[2]{\frac{\mathrm{d}#1}{\mathrm{d}#2}}

\newcommand\antiddots{\mathinner{\mkern2mu\raise1pt\hbox{.}\mkern2mu
    \newline \raise4pt\hbox{.}\mkern2mu\raise7pt\hbox{.}\mkern1mu}}

\def\levy{\mathcal{L}}

\def\IDoS{\mathcal{N}}
\def\kstar{k_c}
\def\thetastar{\theta_c}
\def\CumPos{\kappa^+}
\def\CumNeg{\kappa^-}
\def\R{\mathrm{R}}
\def\LDF{F}

\def\vb{b}

\begin{document}

\renewcommand{\labelitemi}{$\bullet$}
\renewcommand{\labelitemii}{$\star$}

\selectlanguage{english}

\title{The generalized Lyapunov exponent for the one-dimensional \\ Schr\"odinger equation with Cauchy disorder: some exact results}

\author{Alain Comtet}
\affiliation{LPTMS, Universit\'e Paris Saclay, CNRS, 91405 Orsay, France}
\author{Christophe Texier}
\affiliation{LPTMS, Universit\'e Paris Saclay, CNRS, 91405 Orsay, France}
\author{Yves Tourigny}
\affiliation{School of Mathematics, University of Bristol, Bristol BS8 1UG, United Kingdom}

\date{June 21, 2022}


%
%




\begin{abstract}
 We consider the one-dimensional Schr\"odinger equation with a random potential and study the cumulant generating function of the logarithm of the wave function $\psi(x)$, known in the literature as the ``\textit{generalized Lyapunov exponent}''; this is tantamount to studying the statistics of the so-called ``\textit{finite size Lyapunov exponent}''. The problem reduces to that of finding the leading eigenvalue of a certain non-random non-self-adjoint linear operator defined on a somewhat unusual space of functions. We focus on the case of Cauchy disorder, for which we derive a secular equation for the generalized Lyapunov exponent. Analytical expressions for the first four cumulants of $\ln|\psi(x)|$ for arbitrary energy and disorder are deduced. In the universal  (weak-disorder/high-energy) regime, we obtain simple asymptotic expressions for the generalized Lyapunov exponent and for all the cumulants. The large deviation function controlling the distribution of $\ln|\psi(x)|$ is also obtained in several limits. As an application, we show that, for a disordered region of size $L$, the distribution $\mathcal{W}_L$ of the conductance $g$ exhibits the power law behaviour $\mathcal{W}_L(g)\sim g^{-1/2}$ as $g\to0$.
\end{abstract}

\maketitle


\section{Introduction}
\label{sec:Intro}

Solvable models of disorder in one dimension have played an important role in the theory of Anderson localization, providing exact results which have improved our understanding of the underlying physical mechanisms.
Some of the milestones in the developement of the theory of one-dimensional localization are~:
the conjecture by Mott and Twose \cite{MotTwo61} that, in the presence of disorder, every eigenstate becomes localized~; 
the first numerical simulations \cite{DeaBac63,Bor63}; 
the proof of the localisation of the high energy states \cite{Bor63}. Following the development of Furstenberg's theory of products of random matrices \cite{Fur63},
a complete proof of the Mott--Twose conjecture was given a few years later in \cite{MatIsh70, Ish73}.
The pure point nature of the spectrum was demonstrated in Refs.~\cite{GolMolPas77,PasFig78}.  

Despite the apparent simplicity of one-dimensional disordered systems, exact results are scarce.
Explicit formulae for the density of states and the localisation length have been obtained only for a few models. In order to explain the scope of the present paper, it is useful to review briefly these solvable models.
\begin{itemize}[label={$\bullet$},leftmargin=*,align=left,itemsep=0.1cm]
\item
\textit{Discrete models~:}
The first exact result can be found in Dyson's seminal paper \cite{Dys53}, where the spectral density for the one-dimensional tight-binding model with random hopping, distributed according to a gamma law, was obtained 
\cite{footnote1}.
Another famous solvable case is the Lloyd model (tight-binding model with a Cauchy-distributed on-site potential), for which the Green's function at coinciding points can be obtained in any dimension \cite{Llo69} (see also \cite{Ish73,Luc92} for a detailed discussion of the one-dimensional case). 
A third type of distribution leading to exact solution for the tight-binding model is the symmetric exponential \cite{BarLuc90}.

\item
\textit{Continuous models with Gaussian white noises~:}
Amongst the continuous models, those where the disorder takes the form of a Gaussian white noise are the most straightforward, as the disorder is then characterised by only two parameters (the mean value and the weight of the two-point correlator).
A simple example is provided by the Schr\"odinger equation, with Hamiltonian $H=-\partial_x^2+V(x)$ 
(the mean value plays no role in this case). 
Its density of states was obtained by Halperin \cite{Hal65} and its localisation length in Refs.~\cite{LifGrePas88,Nie83,DerGar84}.  
Another model relevant in several physical contexts is the Dirac equation with random mass, $\mathcal{H}_\mathrm{D}=\I\sigma_2\,\partial_x+\sigma_1\,m(x)$, where $\sigma_i$ are the Pauli matrices (see \cite{TexHag10} for a brief review)~; here the mean value of the mass cannot be removed.
This model is related to the supersymmetric Schr\"odinger operators $H_\pm=-\partial_x^2+m(x)^2\pm m'(x)$ (the two partners appear in $\mathcal{H}_\mathrm{D}^2$), which plays a central role in the problem of classical diffusion in a random force field (the so-called Sinai problem).
The density of states for this model was found by Erikmann \& Ovchinnikov \cite{OvcEri77} and rediscovered independently  by Bouchaud {\it et al.} \cite{BouComGeoLeD90}, who also worked out the localisation length. 
Those results were reproduced later by a different method in \cite{BalFis97}.

\item 
  \textit{Continuous models with non Gaussian white noises~:}
Exact results can also be obtained for models involving non-Gaussian white noises, when the disorder is modelled by using delta interactions with random uncorrelated positions and random weights. 
A well-known example is the Frisch-Lloyd model \cite{FriLlo60}--- also known as the "liquid alloy" model \cite{Ish73,Nie83})--- which is the Schr\"odinger
equation with a potential of the form $V(x)=\sum_nv_n\delta(x-x_n)$.
Some solvable cases, involving exponentially-distributed weights, were found by Nieuwenhuizen \cite{Nie83,Nie84,ComTexTou10}. The Frisch--Lloyd model was generalized in \cite{ComTexTou10}; in particular, the 
supersymmetric case where the disorder is modelled by a general L\'{e}vy process was discussed in \cite{ComTexTou11} and \cite{ComTexTou13}, and some solvable cases were found.

\item
  \textit{Disorder with spatial correlations~:} 
 The case of disorder with spatial correlation is more difficult to analyze.
 A solution 
 was obtained for the Dirac/supersymmetric Hamiltonian with a mass that takes the form of a random telegraph noise (with exponentially-decaying correlations)~\cite{ComDesMon95}. 
 This result has found an application to spin chain models \cite{GogNerTsvYu97,FabMel97}.
  The lattice model for spatially-correlated Cauchy disorder was studied, showing that the averaged local Green's function is independent of the correlation length~\cite{Koz14}.

\item
  \textit{Mixed (continuous) models~:}
  The fact that the Schr\"odinger and supersymmetric Hamiltonians exhibit very different properties has motivated the study of the mixed Hamiltonian $H=-\partial_x^2+m(x)^2+m'(x)+V(x)$. 
  The case where $V$ and $m$ are two Gaussian white noises (uncorrelated or correlated) was solved in \cite{HagTex08} (see also \cite{ComLucTexTou13}).
  The mixed case where $m(x)$ is a Gaussian white noise and $V(x)$ is a non-Gaussian white noise consisting of delta interactions with positive weights $v_n$ can be used to study absorption in the Sinai problem \cite{TexHag09}; the problem was also analyzed in the $v_n\to\infty$ limit with the Real Space Renormalization Group method 
  in Ref.~\cite{LeD09}. 
It was shown in Ref.~\cite{GraTexTou14} that the problem becomes solvable when the mass disorder strength $g$ and the mean value of the weights satisfy $g=2\overline{v_n}$.
Let us finally mention that certain continuum limits of random $2\times2$ matrices lead to models that combine up to three Gaussian white noises; the scaling forms that the Lyapunov exponent and the density of states can take were classified in Ref.~\cite{ComLucTexTou13}. 
\end{itemize}

So far, our review has been concerned exclusively with analytical results for the density of states and the Lyapunov exponent.
When we look beyond these basic quantities, exact results become very rare indeed.
One can mention the individual energy-level distributions --the extreme value spectral statistics-- for various models \cite{GreMolSud83,Tex00,TexHag10}, or various properties of the Dirac/supersymmetric model at the band center (for instance, wave function correlations and moments \cite{ComTex98,SheTsv98}, the distribution of the transmission probability \cite{SteCheFabGog99}, or the distribution of the Wigner time delay \cite{SteCheFabGog99,Tex99,Tex16}).
Results that are exact in some asymptotic sense are also known for various models (see the reviews \cite{Gog82,LifGrePas88,Luc92} and, for the Sinai model, \cite{LeDMonFis99}).

A conjecture which has generated much discussion amongst researchers working on Anderson localization is that  of \og single parameter scaling \fg{} (SPS). This conjecture, introduced in~\cite{AbrAndLicRam79}, says that distribution functions of physical observables, like the conductance, are effectively controlled by a single parameter~\cite{AndThoAbrFis80,Sha86}.
A proper examination of the validity (or otherwise) of this conjecture requires the study of {\em fluctuations}. It was first investigated from a phenomenological point of view, based on a composition law for the transmission amplitude and an \textit{ad hoc} random phase assumption \cite{AndThoAbrFis80,CohRotSha88}.

The fluctuations may be studied by computing the {\em cumulants}
\begin{equation}
  \label{eq:DefCumulants}
  \gamma_n = \lim_{x\to\infty}\frac{1}{x}\,\mean{\left(\ln|\psi(x)|\right)^n}_c\,.
\end{equation}
Here, $\psi(x)$ is the wave function that solves the initial value-problem associated with the model, and
$\mean{\cdots}$ denotes averaging over the disorder ($\smean{x^2}_c=\smean{x^2}-\smean{x}^2$, etc). 
The existence of the limit \eqref{eq:DefCumulants} is a non-trivial matter but, for models that can be formulated in terms of products of random matrices, one can use the theory expounded in \cite{BouLac85,BenQui16}.
The first cumulant $\gamma_1$ is the Lyapunov exponent~: according to Borland's conjecture \cite{Bor63}, it provides a measure of the localization of the eigenfunctions that solve the (Sturm-Liouville) spectral problem associated with the model.
For the Lloyd model,  
Deych, Lisyanski and Altshuler were able to derive an analytic formula for the variance $\gamma_2$ \cite{DeyLisAlt00,DeyLisAlt01}. 
For the same model, Titov and Schomerus obtained a complicated analytical form for the third cumulant, and weak-disorder estimates for $\gamma_3$, $\gamma_4$ and $\gamma_5$ \cite{TitSch03}.
They also developed a recursive approach for a different model, equivalent to the Schr\"odinger equation $-\psi''(x)+V(x)\psi(x)=E\psi(x)$ with a Gaussian white noise potential (Halperin's model), leading to formulae for the cumulants
in terms of multiple integrals \cite{SchTit02}. 
A simpler representation of the variance $\gamma_2$ was obtained in \cite{RamTex14} for Halperin's model, and also for the Dirac equation with a random mass.
More recently, a representation of the variance in terms of a {\em single} integral,
valid for the Schr\"odinger equation with any random potential exhibiting local correlations,
was derived in \cite{Tex20}--- thus simplifying and extending previous results; this is the formula contained in Eq.~\eqref{eq:Gamma2TexierJSP2020} below.

We stress an important difference between the approaches used in Refs.~\cite{SchTit02,TitSch03} and that used in Refs.~\cite{Tex20,ComTexTou19} (initiated in Ref.~\cite{FyoLeDRosTex18}).
In both cases, the starting point is the formulation of a certain spectral problem in which the leading eigenvalue is precisely the cumulant generating function
\begin{equation}
  \label{eq:DefGLE}
  \Lambda(q) = \lim_{x\to\infty}\frac{1}{x}\,\ln\mean{|\psi(x)|^q} 
  = \sum_{n=1}^\infty \frac{\gamma_n}{n!}\,q^n
\end{equation}
known also as the {\em generalized Lyapunov exponent (GLE)} \cite{footnote2}.  
In Refs.~\cite{SchTit02,TitSch03}, the GLE is computed perturbatively in the parameter $q$, starting from a standard equation of the Fokker-Planck type with stationary solution \cite{footnote3}.  
On the other hand, \cite{FyoLeDRosTex18,Tex20,ComTexTou19,Tex20b} aims at calculating the GLE {\em non-perturbatively} for a finite value of the parameter $q$. This is obviously a much more difficult undertaking; in particular, it requires a careful consideration of the function space in which the eigenfunction (of the non-random transfer operator) corresponding to the GLE should be sought.
The correct choice of ``boundary conditions'', proposed in Ref.~\cite{FyoLeDRosTex18} for a specific model, was identified in great generality in Refs.~\cite{Tex20,ComTexTou19}, by relating the problem
to a certain representation of the group  $\mathrm{SL}(2,\mathbb{R})$. The relevance of group-theoretical
considerations in this context comes from the fact that, in
the disordered models considered, the solution of the initial-value problem can be expressed in terms of a product of random matrices in $\mathrm{SL}(2,\mathbb{R})$ \cite{ComTexTou10,ComTexTou11,ComTexTou13}.

Despite this progress, the computation of the GLE remains, in general, an extremely difficult problem and,
to the best of our knowledge--- apart from a somewhat trivial case corresponding to a product of triangular  $2\times2$  matrices (cf. Subsection~7.4 of Ref.~\cite{Tex20})--- no tractable model has yet been found.  
The aim of this article is to present and analyze a model which is ``almost solvable'', in the sense that one can write down explicitly a secular equation for the generalised Lyapunov exponent \cite{footnote4}.


As in our previous works \cite{Tex20,ComTexTou19}, the present paper addresses the problem of computing the GLE, identified as the leading eigenvalue of a certain non-random, non-self-adjoint linear operator. This spectral problem was precisely stated in Refs.~\cite{Tex20,ComTexTou19} in the more general context of products of random $2 \times 2$ matrices.
The general formalism introduced in these two papers was applied to several models for which the spectral problem remained unsolvable, hence we proceeded through a perturbative approach in the conjugated parameter $q$~: 
the two first terms of the expansion in powers of $q$ were obtained, providing access to the growth rate and to the variance of the logarithm of the matrix products.
More recently, models with power law disorder were studied by one of us \cite{Tex20b}, where the GLE was studied by a different perturbative approach, in the disorder strength, leading to a weak disorder expression of the GLE for those models. 
The essential novelty of the present paper is that, by focusing exclusively on the case of Cauchy disorder, and by exploiting its special features, we are able to obtain results that are {\it non-perturbative} in both the parameter $q$ and the disorder strength. Hence we get much more information on the large deviations of the wave function. At present, we know of no other model for which such a thorough analysis is feasible.


\subsection{Some physical motivations for Cauchy disorder}

Amongst the models with Cauchy disorder, the Lloyd model, i.e. the one-dimensional tight-binding model 
$
   -\psi_{n+1}+V_n\,\psi_n-\psi_{n-1} = \varepsilon\,\psi_n\,,
$
with potentials $V_n$ that are independent and identically distributed (i.i.d.) with probability density $P(V)=(\vb/\pi)/\big[V^2+\vb^2\big]$, has received the most attention.
As mentioned earlier, from the theoretical point of view, this model is remarkable for the fact that
the Lyapunov exponent and the density of states are relatively easy to compute \cite{Ish73,Luc92,DeyLisAlt00}. On the hand, from the physical point of view, the model is somewhat pathological since 
the second moment $\smean{V_n^2}$ is infinite. 
Nevertheless it is relevant in various contexts.

The first is the quantum kicked rotor, a model exhibiting the phenomenon of dynamical localization (localization of the wave function in momentum space)~: the Floquet eigenstates in momentum space can be shown to obey the same equation as the wave functions of the tight-binding model with Cauchy disorder 
\cite{FisGrePra82,GrePraFis84}.

The second application appeared recently in connection with various models of disordered ladders. 
Even though the disorder in the ladder has finite moments $\smean{V_n^2}<\infty$, it was shown in Ref.~\cite{Luc19} that,  when the ladder is such that its spectrum exhibits a flat band, the problem can be mapped onto the one-dimensional tight-binding model with an effective potential $V_n^\mathrm{eff}$ distributed according to the Cauchy law, i.e. such that $\smean{(V_n^\mathrm{eff})^2}=\infty$.

Finally, the solvability of the Lloyd model has been used in \cite{Mon18} in order to determine the various topological phases of a disordered Kitaev chain.

\subsection{A continuous model}
\label{subsec:ContinousModel}

Instead of considering lattice models, we find it convenient to work here with a continuous model.  
We start with the Schr\"odinger equation
\begin{equation}
  \label{eq:Schrodinger}
  -\psi''(x) + V(x)\,\psi(x) = E\,\psi(x)
\end{equation}
with a potential 
\begin{equation}
  \label{eq:DeltaPotential}
  V(x)=\sum_n v_n\,\delta(x-x_n)
\end{equation}
consisting of impurities distributed along the real line at random uncorrelated positions $x_n$ with a mean uniform density~$\rho$. In this expression, the weights $v_n$ are drawn independently from some distribution whose probability density function is denoted $p(v)$.

Under these assumptions, the integral of the potential
$W(x)=\int_0^x\D t\,V(t)$ is a so-called compound Poisson process, and so we can introduce the L\'evy exponent $\levy(s)$ \cite{App04,ComTexTou11,GraTexTou14}, defined implicitly by 
\begin{equation}
  \label{eq:DefLevy}
  \mean{\EXP{-\I s\int_0^x\D t\,V(t)}} = \EXP{-x\,\levy(s)}\,.
\end{equation}
Equivalently, the generating functional of the disordered potential takes the form~\cite{GraTexTou14} 
\begin{equation}
  G[h] \eqdef \smean{\EXP{-\I\int\D x\,h(x)\,V(x)}}
  =\EXP{- \int\D x\,\levy(h(x))}
  \,.
\end{equation}
The L\'evy exponent is given explictly by
\begin{equation}
  \label{eq:Levy}
  \levy(s) = \rho\,\left[ 1 - \hat p(s) \right]
  \hspace{0.5cm}\mbox{where }
  \hat p(s) = \int\D v\, p(v)\, \EXP{-\I vs}\,.
\end{equation}
In the case where the weights are Cauchy-distributed, we have
$p(v)=(\vb/\pi)/\big[v^2+\vb^2\big]$, and so 
\begin{equation}
  \levy(s)=\rho\,\big(1-\EXP{-\vb\,|s|}\big)\,.
\end{equation}
The resulting model is still too difficult to solve.
However, a simplification occurs by considering the high-density limit with vanishing weights 
\begin{equation}
  \rho\to\infty 
  \hspace{0.25cm} \mbox{and} \hspace{0.25cm}  
  \vb\to0 
  \hspace{0.25cm} \mbox{with} \hspace{0.25cm} 
  \rho \vb=c
  \hspace{0.25cm} \mbox{fixed.} 
\end{equation}
In this limit, the L\'evy exponent becomes
\begin{equation}
  \label{eq:LevyCauchy}
  \levy(s) = c\, |s|\,.
\end{equation}
It is the L\'{e}vy exponent associated with the so-called $\alpha$-stable L\'evy process $W(x)=\int_0^x\D t\,V(t)$ (with $\alpha=1$). In other words, in this limit, $W(x)$ is distributed according to the Cauchy law
\begin{equation}
  p_x(W) = \frac{c\,x/\pi}{W^2+(c\,x)^2}
  \:.
\end{equation}
The parameter $c$ is the strength of the disorder.
This is the model studied here.
This continuous model differs from the discrete (tight-binding and Kronig-Penney) models with Cauchy disorder considered in Refs.~\cite{Ish73,Luc92,DeyLisAlt00,DeyLisAlt01,TitSch03}. Nevertheless, they all exhibit similar properties in the universal high-energy/weak-disorder regime (i.e. at the band edge of the lattice model with vanishing disorder).

\subsection{Main results}

One key result of the paper is the derivation of the following secular equation for the generalized Lyapunov exponent $\Lambda(q)$~:
\begin{equation*}
     \kstar^{q+1} 
  \frac{\Gamma\Big(\frac{\I\Lambda}{2\kstar} + \frac{q}{2} +1 \Big)}
       {\Gamma\Big(\frac{\I\Lambda}{2\kstar} - \frac{q}{2}\Big)}
  =
  \left( \kstar^* \right)^{q+1}
  \frac{\Gamma\Big(\frac{-\I\Lambda}{2\kstar^*} + \frac{q}{2} +1 \Big)}
       {\Gamma\Big(\frac{-\I\Lambda}{2\kstar^*} - \frac{q}{2}\Big)}
\end{equation*}
where $\kstar^2=E+\I c$ combines the energy $E$ and the disorder strength $c$.
Several exact results are deduced from its analysis.
First, we derive analytical expressions for the first four cumulants, 
valid for arbitrary energy and disorder strength; see Eqs.~(\ref{eq:Gamma1Cauchy},\ref{eq:Gamma2Cauchy},\ref{eq:Gamma3Cauchy},\ref{eq:Gamma4Cauchy}).
Turning then to the high-energy limit, we obtain the following expression for the GLE:
\begin{equation*}
  \lim_{E/c\to+\infty}
  \frac{\Lambda(q)}{\gamma_1}
   = \frac{2}{\pi}\,(1+q)\,\tan\left(\frac{\pi q}{2}\right)
\end{equation*}
for $q\in]-3,1[$. 
We also find expressions for all the cumulants:
for $n$ even and $E\gg c$, one has
\begin{equation*}
  \gamma_n\simeq4\pi^{n-2}(2^n-1)|B_n|\,\gamma_1
  \hspace{0.5cm}\mbox{and }
  \gamma_{n-1}\simeq\gamma_n/n
\end{equation*}
where the $B_n$ are the Bernoulli numbers and $\gamma_1\simeq c/(2\sqrt{E})$ is the Lyapunov exponent. 
The divergence of the GLE, i.e. of the moments $\smean{|\psi(x)|^q}$, for $q\to1^-$ and $q\to-3^+$ is characteristic of power-law disorder. 
Using a Legendre transform of $\Lambda(q)$, we then deduce the large deviation function controlling the distribution of $\ln|\psi(x)|$.
We show that the distribution of the conductance for a disordered region of size $L$ exhibits the power-law singular behaviour
\begin{equation*}
   \mathcal{W}_L(g) \underset{g\to0}{\sim} g^{-1/2}
\end{equation*}
This is in agreement with recent numerical simulations~\cite{MenMarGopVar16}.
The high-energy/weak-disorder results obtained in the paper are expected to be \textit{universally valid} for every model with a disordered potential characterised by the power-law tail $p(V)\sim V^{-2}$.
Finally, we also derive various results away from the universal regime.

\subsection{Outline}

Section~\ref{sec:GLE} gives a brief and simplified presentation of the general formalism of Ref.~\cite{Tex20}, adapted to the specific case considered here.
We characterize the GLE $\Lambda(q)$ as the leading eigenvalue of a non-random linear operator.
As a warm-up exercice,  in Section~\ref{sec:PerturbationInQ}, we use the perturbative approach (in powers of $q$) described in Refs.~\cite{Tex20,ComTexTou19}.
In Section~\ref{sec:SecularEquation}, we go beyond the perturbative analysis and obtain the secular equation satisfied by the GLE for fixed values of $q$.
The exact analytical expressions for the first four cumulants are deduced in Section~\ref{sec:FFC}.
The universal (weak-disorder) regime is discussed in Section~\ref{sec:UniversalRegime}. The zero-energy limit and the limit of large negative energy are studied in Section~\ref{sec:BeyondUniversal}.
The GLE is a particular eigenvalue of a spectral problem~: the full spectrum of eigenvalues is discussed in Section~\ref{sec:FullSpectrum}. 
The study of the wave function fluctuations and of the distribution of the conductance is carried out in Section~\ref{sec:LargeDev}.
Finally, in Section~\ref{sec:Alain} we indicate briefly how our somewhat unusual spectral problem relates to some recent works on  the 
spectral problem for non-self-adjoint generalizations of the Schr\"odinger equation with a Coulomb potential.



\section{The GLE as the leading eigenvalue of a non-random linear operator}
\label{sec:GLE}

In this section, we recall the main formalism introduced in Refs.~\cite{Tex20,ComTexTou19} to study the generalized Lyapunov exponent (GLE) of products of random matrices in the group $\mathrm{SL}(2,\mathbb{R})$. 
We focus here on a specific case, which leads to a simple derivation of the main equation for the spectral problem.

\subsection{Formulation in terms of a product of random matrices}


As is well-known from elementary quantum mechanics, for the Kronig--Penney potential \eqref{eq:DeltaPotential}, the solution of the initial-value problem for the Schr\"odinger equation \eqref{eq:Schrodinger} on the half-line $x>0$ can be expressed as a product of random transfer matrices. More precisely, if
we set $E=k^2$ and
recast the equation as a first-order system of two equations for the unknown vector $\big(\psi'(x)\,,\,k\,\psi(x)\big)^\mathrm{T}$ then, by considering 
the equation in each of the intervals $[x_n,x_{n+1}[$, we see that the effect of the delta potential and of the free evolution corresponds  to multiplication by the matrices 
\begin{equation}
  \label{eq:MatricesOfInterest}
  N(u_n)=
  \begin{pmatrix}
     1 & u_n \\ 0 & 1
  \end{pmatrix}
  \;\;\mbox{and}\;\;
  K(\theta_n)=
  \begin{pmatrix}
    \cos \theta_n & -\sin\theta_n
    \\
    \sin \theta_n & \phantom{-}\cos\theta_n
  \end{pmatrix}
\end{equation}
respectively. 
Both these $2 \times 2$ matrices have unit determinant. Hence they, and any repeated product of them, belong to the group $\text{SL}(2,\mathbb{R})$.
(In the case $E=-k^2$, the matrix $K$ must be replaced by another matrix in $\text{SL}(2,\mathbb{R})$ with entries involving the hyperbolic functions; see \cite{ComTexTou10,RamTex14}.)
The angle of rotation is proportional to the interval length: $\theta_n=k\ell_n$, with $\ell_n=x_{n+1}-x_n$, and the upper off-diagonal coefficient in the matrix $N$ is proportional to the weight of the impurity at $x_n$: $u_n=v_n/k$. In our particular model, the number of impurities in the interval $[0,x]$ is a Poisson process, say $\mathscr{N}(x)$, of intensity $\rho$, i.e. $\mathrm{Proba}\{\mathscr{N}(x)=n\}=\EXP{-\rho x}(\rho x)^n/n!$. 
Therefore, the $\ell_n$ are independent and exponentially distributed with mean $1/\rho$, so that $\mathrm{Proba}\{\ell_n>\ell\}=\EXP{-\rho\ell}$. 
Introducing the random matrices $M_n=K(\theta_n)N(u_n)$, the $q$-th moment of the wave function $\psi$ that solves the initial-value problem can then be written as 
\begin{equation}
  \label{eq:RMP}
  \mean{|\psi(x)|^q} \sim \mean{||\Pi_{\mathscr{N}(x)}\vec{x}_0||^q}
\end{equation}
where
\begin{equation}
  \Pi_n = M_n\cdots M_2M_1
\end{equation}
$||\vec{x}||$ is the usual Euclidean vector norm.
$\vec{x}_0$ is a vector of unit length on which acts the product $\Pi_n$,
expressing the initial conditions. 
For example, $\vec{x}_0=(1\,,\,0)^\mathrm{T}$ corresponds to imposing the initial conditions $\psi'(0)=1$ and $\psi(0)=0$.

\subsection{A spectral problem}
\label{subsec:SPwithRiccati}
A general formalism for the study of the moments $\mean{||\Pi_N\vec{x}_0||^q}$, when $N$ is non-random, was developed in \cite{ComTexTou19, Tex20}, and the necessary adjustments that are needed to cater for the case \eqref{eq:RMP}, where $N={\mathscr N}(x)$ is random, were indicated in \cite{Tex20}. In what follows, we provide a simplified description of this formalism, adapted to our particular class of models.

The matrices in $\text{SL}(2,\mathbb{R})$ act on $\mathbb{R}^2$ by multiplication, and we have expressed the wave function in terms of a matrix acting on some initial vector of unit length. Every vector of unit length can be identified with a {\em direction}, which can be parametrized either in terms of the angle, say $\theta \in [0,\pi)$, it makes with the horizontal axis, or else by the ratio, say $z \in \mathbb{R}$, of the Cartesian coordinates of the vector; in the first case, we speak of the {\em projective semi-circle}, and in the second of the {\em projective line}. 
In the latter parametrization, the action of a matrix $M$ on the projective line corresponds to a M\"obius map \cite{BouLac85}~:
\begin{equation}
   z \mapsto \mathcal{M}(z) = \frac{a\,z+b}{c\,z+d}
  \hspace{0.5cm}\mbox{for }
  M=\begin{pmatrix}
    a & b \\ c & d 
  \end{pmatrix}
  \:.
\end{equation}
Instead of following the vector $(\psi'(x),k\,\psi(x))^\mathrm{T}$,  it will be simpler to track the Riccati variable $z(x)=\psi'(x)/\psi(x)$. 
Accordingly, we define
\begin{align}
  \mathcal{P}_x(z|z_0;q) 
  &\eqdef
  \mean{ \delta(z-z(x)) \,|\psi(x)|^q }
  \\
  \label{eq:PwithExp}
  &=  \mean{ \delta(z-z(x)) \,\EXP{q\int_0^x\D t\,z(t)} }
\end{align}
which determines the moments 
\begin{equation}
  \label{eq:MomProp}
   \mean{|\psi(x)|^q}  = \int\D z\, \mathcal{P}_x(z|z_0;q)
   \:.
\end{equation} 
How it evolves during an infinitesimal interval of ``time'' $[x, x+\D x]$ can be deduced from the following considerations~: 
\begin{enumerate}[label={(\roman*)},leftmargin=*,align=left,itemsep=0.1cm]
\item
  If 
  $[x,x+\D x]$ contains exactly one impurity, then
  the Riccati variable makes a jump $z(x_n^+)=z(x_n^-)+v_n$.
  This occurs with probability $\rho\,\D x$.

\item 
  If the interval $[x,x+\D x]$ contains no impurity, then the evolution of $\psi$ --- and hence also of $z$--- is free. In that case,
  $z(x+\D x)\simeq z(x)-\big(E+z(x)^2\big)\,\D x$ so that, if we introduce
  $\tilde z=z+(E+z^2)\,\D x$, we can write
  $\D z\,\mathcal{P}_{x+\D x}(z|z_0;q)=\D\tilde z\,\mathcal{P}_{x}(\tilde z|z_0;q)$.
  This scenario occurs with probability $1-\rho\,\D x$. 

\item 
  During the interval $[x,x+dx]$, the exponential in \eqref{eq:PwithExp} grows by a factor $1+q\,z\,\D x$.

\item 
  The probability that $[x,x+\D x]$ contains two or more impurities is $o(\D x)$ and may 
  be neglected.
\end{enumerate}
As a result
\begin{align}
  &\mathcal{P}_{x+\D x}(z|z_0;q)
  \simeq    \rho\,\D x\, \mean{ \mathcal{P}_{x}(z-v|z_0;q) }_{v} 
  + \left(1-\rho\,\D x\right)
  \nonumber\\
 &  \times     
       \left( 1+ 2z\,\D x\right)\,\mathcal{P}_{x}\big(z+[E+z^2]\,\D x|z_0;q\big)
     \left( 1+ q\,z\,\D x\right)
\end{align}
where $\mean{\cdots}_v$ denotes averaging over the impurity weights with distribution $p(v)$.
This leads to 
\begin{align}
  \label{eq:EqForPropagator}
  &\partial_x\mathcal{P}_{x}(z|z_0;q) = \mathscr{L}_q \mathcal{P}_{x}(z|z_0;q)
\end{align}
where $\mathscr{L}_q$ is a non-random linear operator, defined by
\begin{align}  
\label{eq:DefLq}
&\mathscr{L}_q\phi(z) 
  \eqdef \left[ 
       \partial_z(E+z^2) + q\, z
       + \rho \left( \mean{\EXP{- v\,\partial_z }}_v  -1 \right) 
      \right] \phi(z)
\nonumber\\
      &=
       \left[ \partial_z (E+z^2) + q\, z\right]\phi(z)
       + \rho \left[ \mean{\phi(z-v)}_v  -\phi(z)\right]
       \:.
\end{align}
This operator acts in a certain $q$-dependent space of functions whose properties are discussed below.
Eqs.~(\ref{eq:EqForPropagator},\ref{eq:DefLq}) agree with Eq.~(6.11) of Ref.~\cite{Tex20}, specialised to our particular case. 
Eq.~\eqref{eq:EqForPropagator} makes clear that the large-$x$ behaviour of $\mathcal{P}_{x}(z|z_0;q)$ can be obtained from a spectral analysis of the operator $\mathscr{L}_q$~: 
if we assume a discrete spectrum, denoted by $\{\Lambda_n(q)\}_{n \in {\mathbb Z}}$, where  $\Lambda_0(q)$ is the leading eigenvalue, we expect the behaviour  $\mathcal{P}_{x}(z|z_0;q)\sim\EXP{x\Lambda_0(q)}$. 
From \eqref {eq:MomProp}, we then deduce that
\begin{equation}
  \label{eq:SpectralMethodToGetTheGLE}
  \Lambda(q) = \Lambda_0(q)
  \:,
\end{equation}
so that the GLE can be obtained from the spectral analysis of $\mathscr{L}_q$.
This approach relies on
(i) a precise definition of the underlying spectral problem, and in particular of the nature of the space of functions on which the operator $\mathscr{L}_q$ acts~;
(ii) the existence of a spectral gap--- that is, the discrete spectrum must be such 
that the leading eigenvalue $\Lambda_0(q)$ is isolated from the rest of the spectrum. 
We proceed to discuss these two important points.

The case $q=0$ has been studied extensively in the literature pertaining to products of random matrices, and is well-understood. $\mathcal{P}_{x}(z|z_0;0)$ is the distribution of the process $z(x)$ and, under broad conditions, it has a limit law
\cite{footnote5}~:  
$\mathcal{P}_{x}(z|z_0;0)\to f(z)$ as $x\to\infty$, where the stationary probability density satisfies
$\mathscr{L}_0f(z)=0$ or, more explicitly,
\begin{equation}
  \label{eq:DysonSchmidt}
  \partial_z\big[(E+z^2)f(z)\big] + \rho\,\big[\mean{f(z-v)}_v-f(z)\big]=0\,.
\end{equation}
In this case, $f$ is a right-eigenfunction corresponding to the eigenvalue $0$ and, under the same broad conditions, it may be shown that it is the leading eigenvalue, so that $\Lambda(0)=0$.
The integro-differential equation \eqref{eq:DysonSchmidt} is the form taken
by the so-called Dyson-Schmidt equation for our particular model.
This equation makes clear that the stationary density $f$ exhibits the asymptotic behaviour
\begin{equation}
  \label{eq:riceFormula}
  \lim_{z\to-\infty} (-z)^{2}\,f(z)   = \lim_{z\to+\infty} z^{2}\,f(z)\,.
\end{equation}
It expresses that the probability current associated with the stationary distribution takes the same value  at $+\infty$ and $-\infty$. 
The current, i.e. the value of the limit \eqref{eq:riceFormula}, coincides with the integrated density of states (IDoS) $\IDoS(E)$  of the random Schr\"odinger operator \cite{LifGrePas88,Luc92,ComTexTou10}~: this is the well-known {Rice formula} \cite{Kot76}.
We proceed to argue that a relation of the type \eqref{eq:riceFormula}, suitably generalised for $q \ne 0$,
provides the correct auxiliary condition
that must be imposed on {\em every} right-eigenfunction of $\mathscr{L}_q$ in order to complete the definition of the spectral problem.

For arbitrary $q$, but small density $\rho$ (or small weights $v_n$), the operator $\mathscr{L}_q$ can be viewed as a perturbation of the differential operator $\mathscr{D}_K(q)=\partial_z(k^2+z^2) + q\, z$ and so, it is at least plausible that the functional setting should be the same for both problems. 
Now, the spectral problem for $\mathscr{D}_K(q)$ takes a particularly simple form if, instead of working on the projective line, we go over to the projective semi-circle, which uses 
$\theta=\mathrm{arccotg}(z/k)$ as the independent variable. 
In this alternative parametrization, $\mathscr{D}_K(q)$ is transformed into 
$\widetilde{\mathscr{D}}_K(q)= -\partial_\theta+q\,\cotg\theta$
and the natural domain for this operator is the space of $\pi$-periodic functions. It follows easily (see \cite{Tex20}, Appendix B) that it has a discrete spectrum consisting of the eigenvalues 
$\Lambda^{(0)}_n(q)=-2\I n\,k$ (with $n\in\mathbb{Z}$), with corresponding right-eigenfunctions
$\EXP{2\I n\theta}\,(\sin\theta)^q$. The spectrum does not depend on how we parametrize the space of directions. Expressed as functions on the projective line, the corresponding right-eigenfunctions of $\mathscr{D}_K(q)$ are
\begin{equation}
  \label{eq:EigenvectorDK}
  \varphi_n^\R(z;q) = \frac{1}{\sqrt{\pi}}\,\left(\frac{z+\I\,k}{z-\I\,k}\right)^n\,(z^2+k^2)^{-1-q/2}
\end{equation}
and, together, form a countable basis for the ``correct'' space of functions on the projective line. These basis functions all exhibit the asymptotic behaviour $\varphi_n^\R(z;q)\simeq A\,|z|^{-2-q}$. The key observation is that coefficient of the power law is the same at $z\to+\infty$ and $z\to-\infty$.

These remarks allow us to complete the definition of the spectral problem for
\begin{equation}
  \label{eq:SpectralPb}
  \mathscr{L}_q \phi(z;\Lambda) = \Lambda \,\phi(z;\Lambda)
\end{equation}
by imposing that the two limits
\begin{equation}
  \label{eq:DefApm}
  A_\pm(\Lambda) \eqdef \lim_{z\to\pm\infty} (\pm z)^{2+q}\,\phi(z;\Lambda) 
\end{equation}
must exist and coincide:
\begin{equation}
  \label{eq:BoundaryConditions}
  A_+(\Lambda) = A_-(\Lambda)
  \:.
\end{equation}
This equation is precisely the secular equation satisfied by the eigenvalues of the non-self-adjoint operator~$\mathscr{L}_q$.

We close the paragraph with several remarks~:
\begin{itemize}[label={$\bullet$},leftmargin=*,align=left,itemsep=0.1cm]
\item
The condition \eqref{eq:BoundaryConditions} is dictated by representation-theoretical considerations whose
relevance to the correct formulation of the spectral problem was an important outcome of~\cite{Tex20,ComTexTou19}.
For products of random matrices in $\mathrm{SL}(2,\mathbb{R})$, the spectral problem involves a non-random operator which depends ---in a usually very complicated way--- on $\mathscr{D}_K(q)$ and on two other operators which, together, span a representation of the Lie algebra of $\mathrm{SL}(2,\mathbb{R})$. 
The operators in this Lie algebra act on functions defined on the projective line and, roughly speaking, condition of the type (\ref{eq:DefApm},\ref{eq:BoundaryConditions}) characterises the function spaces associated with a particularly important family, indexed by~$q$, of irreducible representations of the group;
see for instance Chapter~2 of the book \cite{Unt00}.
In this connection, we remark that the case $q=-1$ corresponds to a unitary representation of the group $\mathrm{SL}(2,\mathbb{R})$~; the eigenfunctions \eqref{eq:EigenvectorDK} for this case appear in~\cite{Itz69}.

\item 
Since the operator ${\mathscr L}_q$ is not self-adjoint, each eigenvalue is associated with a pair of
right and left-eigenfunctions. Above, we have worked exclusively with the auxiliary condition
satisfied by the right-eigenfunctions.
If we work instead with left-eigenfunctions, we find that they satisfy an 
auxiliary condition like \eqref{eq:DefApm}, but with $q$ replaced by $-q^\ast -2$ \cite{ComTexTou19}. We then obtain a secular equation that is equivalent.

\item
  We will denote by $\Lambda_n(q)$ the solutions of the equation \eqref{eq:BoundaryConditions} and 
  by $\Phi^\R_n(z;q)$ the related right-eigenvectors of $\mathscr{L}_q$, which thus satisfy 
\begin{align}
    \label{eq:RightEV}
    &\left[ \partial_z (E+z^2) + q\, z \right]\Phi^\R_n(z;q)
   \\\nonumber
   + &\rho\left[ \mean{\Phi^\R_n(z-v;q)}_v  - \Phi^\R_n(z;q)\right]  
    = \Lambda_n(q)\, \Phi^\R_n(z;q)
    \:.
\end{align}

\item
For products of random matrices, the existence of a spectral gap below the leading eigenvalue, when $q$ is small, has been investigated; see Ref.~\cite{BouLac85}, Chapter V.
For Cauchy disorder, the spectral gap is discussed in Section~\ref{sec:FullSpectrum} below.

\item
For the specific case of a Gaussian white noise potential, Equation \eqref{eq:SpectralPb} reduces to a differential equation~: it suffices to take the limit $\rho\to\infty$ and $v_n\to0$ with $\mean{v_n}=0$ and  $\sigma=\rho\mean{v_n^2}$ fixed, which yields $\rho\left[ \mean{\Phi^\R_n(z-v;q)}_v  - \Phi^\R_n(z;q)\right] \to (\sigma/2)\partial_z^2\Phi^\R_n(z;q)$ in \eqref{eq:RightEV}, leading to Eq.~(105) of Ref.~\cite{FyoLeDRosTex18}. 
In that reference, Equation \eqref{eq:BoundaryConditions} was used to determine the GLE numerically
to a great accuracy.
\end{itemize}

\subsection{The spectral problem in Fourier space~: from an integro-differential to a differential operator}
\label{subsec:SPwithFourier}

Unless the integro-differential equation \eqref{eq:SpectralPb} can be solved explicitly for $\phi(z;\Lambda)$, Eq.~\eqref{eq:BoundaryConditions} appears to be of limited use at this stage. 
The problem becomes more tractable if we look instead for the Fourier transform~\cite{footnote6bis}
\begin{equation}
  \widehat{\phi}(s;\Lambda) = \int_{-\infty}^{+\infty}\D z\, \EXP{-\I sz} \,\phi(z;\Lambda)
  \:.
\end{equation}
In Fourier space, Eq.~\eqref{eq:SpectralPb} takes the form
\begin{equation}
  \label{eq:SpectralPbFourier}
  \I s \left[ -\deriv{^2}{s^2}+\frac{q}{s}\deriv{}{s} + E - \frac{\levy(s)}{\I s} \right]   
  \widehat{\phi}(s;\Lambda)
  = \Lambda   \,
  \widehat{\phi}(s;\Lambda)
\end{equation}
where the L\'evy exponent $\levy(s)$ was defined 
by Eqs.~(\ref{eq:DefLevy},\ref{eq:Levy}).

To proceed, we must explain how, knowing the Fourier transform $\widehat{\phi}(s;\Lambda)$, we can make concrete use of the secular equation \eqref{eq:BoundaryConditions} to determine the spectrum. 
The explanation consists of two parts. First, we recall that, for an arbitrary $\Lambda$, the function $\phi(z;\Lambda)$ denotes any non-zero solution of Eq.~\eqref{eq:SpectralPb} such that the limits \eqref{eq:DefApm} 
exist. 
From the existence of these limits, we deduce the following
\begin{enumerate}[label={(\arabic*)},leftmargin=*,align=left,itemsep=0.1cm]
\item 
  For $q > -3/2$, $\phi(z;\Lambda)$ is square-integrable, and so $\widehat{\phi}(s;\Lambda)$ must decay at $\pm \infty$. 

  \item Unless the limits $A_{\pm} (\Lambda)$ vanish, $\phi(z;\Lambda)$ exhibits an algebraic behaviour at infinity. This implies that $\widehat{\phi}(s;\Lambda)$ is not smooth at $s=0$. 
\end{enumerate} 
In order to analyze the local behaviour of $\widehat{\phi}(s;\Lambda)$ at $s=0$, it will be convenient to consider in the first instance the case $q>-1$, so that 
$\phi(z,\Lambda)$ is integrable and its Fourier transform continuous. 
There is then no loss of generality in assuming that $\widehat{\phi} (0;\Lambda) =1$, and we can write
\begin{equation}
\widehat{\phi}(s;\Lambda) 
= \begin{cases}
\phi_-(s;\Lambda) & \text{if $s<0$} \\
\phi_+(s;\Lambda) & \text{if $s>0$}
\end{cases}
\end{equation}
where $\phi_\pm(s;\Lambda)$ is the solution of
\begin{equation}
  \label{eq:ZeDifferentialEquation}
  \left[ -\deriv{^2}{s^2}+\frac{q}{s}\deriv{}{s} + E - \frac{\levy(s)+\Lambda}{\I s} \right] 
  \phi_\pm(s;\Lambda) = 0
  \,,\;\;
  \pm s > 0
  \,,
\end{equation}
that satisfies $\phi_{\pm}(0;\Lambda) = 1$ and vanishes at $\pm\infty$.
Consider first the ``$+$'' equation: in the limit $s \to 0^+$,
$\levy(s)/s$ has a limit
\cite{footnote6} 
and so three terms dominate in the differential equation, namely
$-\phi_+'' + (q/s)\,{\phi}'_+ + (\I\Lambda/s)\,\phi_+ \simeq0$. 
Retaining only the last two 
leads to the behaviour $\phi_+(s;\Lambda)\simeq1-(\I\Lambda/q)\,s$, 
while retaining the first two 
produces a non analytic contribution $s^{q+1}$. 
Thus, for $|\re{q}|<1$, the first terms of the small $s$ 
expansion are 
\begin{equation}
  \label{eq:omegaPlus}
  \phi_+(s;\Lambda) = 1 - \frac{\I\Lambda}{q}\,s + \Omega_+(\Lambda)\,s^{q+1} + \cdots
  \hspace{0.5cm}\mbox{for }s\to0^+
  \:.
\end{equation}
Similar considerations hold for the solution $\phi_-$; thus 
\begin{equation}
\label{eq:omegaMinus}
  \phi_-(s;\Lambda) = 1 - \frac{\I\Lambda}{q}\,s + \Omega_-(\Lambda)\,(-s)^{q+1} + \cdots
  \hspace{0.5cm}\mbox{for }s\to0^-
  \:.
\end{equation}

We arrive at the second part of our explanation. We can write
\begin{equation}
  \label{eq:SplitPhiHat}
  \widehat{\phi}(s;\Lambda)
  = \int\D z\, \phi(z;\Lambda)  - \int\D z\, \phi(z;\Lambda)  \left(1-\EXP{-\I sz}\right)
\end{equation}
Let us now suppose that $\Lambda$ is an eigenvalue, so that Equation \eqref{eq:BoundaryConditions} holds and the limits are {\em equal}.
In the limit $s\to0$, the second integral is then dominated by the tail of the function 
$\phi(z;\Lambda) \simeq A\,|z|^{-2-q}$ for $z\to\infty$, with $A\equiv A_+(\Lambda)=A_-(\Lambda)$. 
More precisely,  using the result in Appendix~\ref{app:FourierReciprocity}, we can relate the power law decay for $z\to\pm\infty$ to the $s\to0$ behaviour 
\begin{equation}
  \label{eq:limitPhiHat}
   \widehat{\phi}(s;\Lambda)
   \underset{s\to0}{\simeq} 
    1 + \alpha_1 \,s - 2A\,\Gamma(-1-q)\,\sin\left(\frac{\pi q}{2}\right)\,|s|^{q+1}
\end{equation}
for $-1<\re(q)<1$, where $\alpha_1$ is some coefficient.
Comparing this with Equations (\ref{eq:omegaPlus},\ref{eq:omegaMinus}), 
we deduce that the secular equation \eqref{eq:BoundaryConditions} for the eigenvalues can be expressed in the equivalent form
\begin{equation}
  \label{eq:GeneralSecularEq}
  \Omega_+(\Lambda)=\Omega_-(\Lambda)\,.
\end{equation}
We expect this to remain true for a larger interval of the parameter $q$ than that assumed here, so that, in particular, the Fourier transform of the right-eigenfunction presents the small-$s$ behaviour
\begin{equation}
  \label{eq:LimitingBehaviourPhiR}
  \widehat{\Phi}^\R_n(s;q)
  \underset{s\to0}{=} 
  \underbrace{ 
     1 - \frac{\I\,\Lambda_n(q)}{q}\,s + \mathcal{O}(s^2) 
  }_{\mathrm{analytic}}
  +  
  \underbrace{ 
     \omega_q\,|s|^{q+1} + \mathcal{O}(s^{q+2}) 
  }_{\mathrm{non\ analytic}}
\end{equation}
where $\omega_q=\Omega_\pm(\Lambda_n(q))$.

Since $\Lambda(q) = \Lambda_0 (q)$, the case $n=0$ will be our prime concern in what follows, and calls for a few remarks:
\begin{itemize}[label={$\bullet$},leftmargin=*,align=left,itemsep=0.1cm]
\item
  For $q$ real, from the definition \eqref{eq:DefGLE}, the leading eigenvalue $\Lambda_0(q)=\Lambda(q)$ must be real.
  As a result, the right-eigenfunction $\Phi^\R_0(z;q)$ is real, and hence also the coefficient $\omega_q$; see Appendix C of Ref.~\cite{Tex20} where the behaviour \eqref{eq:LimitingBehaviourPhiR} was derived.
  
\item
Equation \eqref{eq:SpectralPbFourier} reduces in the case $\Lambda=q=0$ to the equation for the Fourier transform $\hat{f}(s)$
of the invariant probability density $f(z)$, the normalized solution of the Dyson--Schmidt equation \eqref{eq:DysonSchmidt}. After dividing both sides by $s$, the equation takes the form
\begin{equation}
  \label{eq:EqForInvariantDensity}
  \left[ -\deriv{^2}{s^2}+ E - \frac{\levy(s)}{\I s} \right] \hat{f}(s)
  = 2\pi\,\IDoS (E)\,\delta(s)
  \:,
\end{equation}
where $\IDoS(E)$ is the IDoS mentioned earlier in connection with the equality \eqref{eq:riceFormula} \cite{Kot76,ComTexTou10,GraTexTou14,Tex20}.
It is clear from the equation that $\hat{f}(-s)=\hat{f}(s)^*$
and $\pi\IDoS(E)=-\im\big[\hat{f}'(0^+)\big]=-\omega_0$.  
Furthermore, the first cumulant $\gamma_1 = \Lambda'(0)$ may be expressed as
\begin{equation}
\gamma_1= \dashint\D z\,z\,f(z) = -\im[\hat{f}'(0^+)]\,.
\label{eq:integralFormulaForTheFirstCumulant}
\end{equation}
 By putting these results together, we deduce 
 \begin{equation}
  \label{eq:DerivativeFhat2}
  \I\,\hat{f}'(0^+)=\gamma_1-\I\,\pi\,\IDoS\,.
  \end{equation}
 The quantity on the right-hand side is sometimes referred to as the ``complex Lyapunov exponent'' or the ``characteristic function''
 associated with the model \cite{Nie83,Luc92,ComTexTou10,ComTexTou11,ComLucTexTou13,GraTexTou14}.

\item
    In Ref.~\cite{Tex20}, the spectral problem \eqref{eq:SpectralPbFourier} was studied by a perturbative approach (in powers of $q$) for a general Schr\"odinger equation with disorder, leading to a compact formula for the variance
\begin{equation}
  \label{eq:Gamma2TexierJSP2020}
     \gamma_2
    =    \int_0^\infty \frac{\D s}{s}\,       
    \re\left[
      \left(  
         2\gamma_1  - \I\, \deriv{}{s}
      \right)
      \hat{f}(s)^2
    \right]
    \:.
\end{equation}
in terms of the Fourier transform of the invariant probability density, the solution of Equation \eqref{eq:EqForInvariantDensity}.

\end{itemize}
The remainder of the paper will be concerned with the application of the formalism developed in this section to the particular model with Cauchy disorder introduced earlier in Subsection~\ref{subsec:ContinousModel}.


\section{Warm-up~: perturbative approach (in $q$) and the cumulants $\gamma_1$ \& $\gamma_2$}
\label{sec:PerturbationInQ}

It is useful to begin by working out the first two cumulants perturbatively ---that is, following the method of \cite{Tex20}, by making use of the formulae (\ref{eq:EqForInvariantDensity}-\ref{eq:Gamma2TexierJSP2020}). 
This will provide a useful check for the results that will be derived subsequently from an analysis of the secular equation.
The strategy is then to solve this differential equation \eqref{eq:EqForInvariantDensity} for $s>0$,  identify the solution vanishing at $+\infty$ and impose $\hat{f}(0)=1$. 
For the case of Cauchy disorder, this is extremely simple.
Eq.~\eqref{eq:EqForInvariantDensity} yields 
\begin{equation}
  \left[ - \deriv{^2}{s^2} + E + \I\,c  \right] 
  \hat{f}(s) = 0
  \hspace{0.5cm}\mbox{for }s>0
  \:.
\end{equation}
The solution is 
\begin{equation}
  \hat{f}(s) = 
    \EXP{-\kstar\,s}
    \hspace{0.5cm}
    \mbox{for }s>0
  \:,
\end{equation}
where 
\begin{equation}
  (\kstar)^2 = E+\I c = |\kstar|^2\,\EXP{\I\thetastar}
  \hspace{0.5cm} \mbox{with }
  \thetastar \in ]0,\pi[
  \:.
\end{equation}
For $s<0$, we use $\hat{f}(s) = \hat{f}(-s)^*$.
From \eqref{eq:DerivativeFhat2}, we deduce
\begin{equation}
  \kstar = \pi\, \IDoS + \I\,\gamma_1\,,
\end{equation}
so that $\kstar$ coincides with the complex Lyapunov exponent associated with the model (\ref{eq:Schrodinger},\ref{eq:LevyCauchy}).
By considering the real and imaginary parts, we arrive at
\begin{align}
  \label{eq:Gamma1Cauchy}
  &\pi\, \IDoS
  = |\kstar|\,\cos\left({\thetastar}/{2}\right)
  =\sqrt{\frac{\sqrt{E^2+c^2}+E}{2}}
  \\
  &\simeq
  \begin{cases}
    \sqrt{E}+\frac{c^2}{8E^{3/2}} & \mbox{for } E\gg c
    \\
    \sqrt{c/2} & \mbox{for } |E|\ll c
    \\
    \frac{c}{2\sqrt{-E}}& \mbox{for } -E\gg c
  \end{cases}
\end{align}
and 
\begin{equation}
  \label{eq:Gamma1Cauchy}
  \gamma_1 
  = |\kstar|\,\sin\left({\thetastar}/{2}\right)
  \:.
\end{equation}
Thus, for Cauchy disorder we have the property 
\begin{equation}
   \gamma_1 (E) = \pi\, \IDoS(-E)
   \:.
\end{equation} 
In particular, we see that the power-law decay $\IDoS\simeq c/\big(2\pi\sqrt{-E}\big)$ of the IDoS in the limit  $E\to-\infty$ is related to the power-law decay $\gamma_1\simeq c/\big(2\sqrt{E}\big)$ of the 
Lyapunov exponent  as $E\to+\infty$.
The slow power-law decay of the IDoS for $E\to-\infty$ shows that Cauchy disorder shifts states to very large negative energies. This contrasts with the case where the disorder has finite moments, where the IDoS exhibits Lifshitz tails that decay exponentially.

We obtain the variance straightforwardly from~\eqref{eq:Gamma2TexierJSP2020}~:
\begin{align}
  &\gamma_2 = -2\pi \IDoS \int_0^\infty\frac{\D s}{s}\,\im\left[\hat{f}(s)\right]
  \\
  &= 2\pi \IDoS \int_0^\infty\D s\,\frac{\sin(2\gamma_1s)}{s}\,\EXP{-2\pi \IDoS s}
  = 2\pi \IDoS \, \mathrm{arccot}\left(\frac{\pi\,\IDoS}{\gamma_1}\right)
  \nonumber
\end{align}
or, equivalently,
\begin{align}
  \label{eq:Gamma2Cauchy}
  &\gamma_2  = |\kstar|\,\thetastar\,\cos\left({\thetastar}/{2}\right)
  \\
  = &\sqrt{2\Big(\sqrt{E^2+c^2}+E\Big)}\:
  \mathrm{arccot}\left(\sqrt{\frac{\sqrt{E^2+c^2}+E}{\sqrt{E^2+c^2}-E}}\right)\,.
  \nonumber
\end{align}
The formulae (\ref{eq:Gamma1Cauchy}) and (\ref{eq:Gamma2Cauchy}) make it easy to compare the limiting behaviours of the first two cumulants
as $E\to+\infty$ (then $\thetastar\to0$).
We get
\begin{equation}
  \label{eq:SPSCauchyWeak}
  \gamma_2 \simeq 2 \gamma_1
  \hspace{0.5cm}\mbox{for }
  E\gg c
  \:.
\end{equation}

We close the section with some remarks~:
\begin{itemize}[label={$\bullet$},leftmargin=*,align=left,itemsep=0.1cm]
\item
The asymptotic behaviour of the IDoS for $E\to-\infty$ and the decay of the Lyapunov exponent for $E\to+\infty$ are quite different from the ones obtained in the more standard case where the second moment of the disordered potential is finite, usually leading to Lifshitz tails and a faster power law decay for the Lyapunov exponent $\gamma_1\sim1/E$ \cite{AntPasSly81,LifGrePas88,Luc92}.
The origin of the unusual power-law decay was already identified in Ref.~\cite{BieTex08} for a power law disorder characterized by $p(v)\sim|v|^{-1-\alpha}$ for $v\to\pm\infty$ 
with $\alpha\in]0,2[$~:
as was made clear in that paper, the increase of $\ln|\psi(x)|$ due to an impurity is $\sim\ln|v_n/\sqrt{E}|$, so that $\ln|\psi(x)|$ obeys a generalized central limit theorem when the second moment of $\ln|v_n|$ is finite.
However, the signature of the power-law disorder can be seen in the energy decay of the Lyapunov exponent, namely $\gamma_1\sim E^{-\alpha/2}$ for $\alpha\in]0,2[$.

\item
The relation \eqref{eq:SPSCauchyWeak} was obtained for the discrete tight-binding model with Cauchy disorder in Refs.~\cite{DeyLisAlt00,DeyLisAlt01,TitSch03}. 
It is a manifestation of the \og single parameter scaling \fg{} (SPS) property, i.e. the fact that the distribution of $\ln|\psi(x)|$  (or the distribution of the conductance of a disordered slice) is controlled by a unique scale.
Cauchy disorder is responsible for an additional factor of $2$ compared to the standard case of disorder with finite second moment. 
A broader perspective was given in Ref.~\cite{Tex20b} where the factor $2$ was related to the exponent of the tail of the disorder distribution $p(v)\sim|v|^{-1-\alpha}$ (or equivalently, the exponent controlling the L\'evy exponent $\levy(s)\sim|s|^\alpha$ for $s\to0$)~: Eq. \eqref{eq:SPSCauchyWeak} then generalizes to $\gamma_2\simeq(2/\alpha)\,\gamma_1$. 
Furthermore, the relation was shown to be a particular case of a more general relation between even and odd cumulants in the case of power-law disorder 
\begin{equation}
  \label{eq:SPSTexier2020}
 \gamma_{2m}\simeq\frac{2m}{\alpha}\,\gamma_{2m-1} 
  \hspace{0.5cm}\mbox{for }
  E\gg\:\mathrm{disorder\ strength}
\end{equation}
with $m\in\mathbb{N}^*$.
The model that we study in this paper corresponds to the case $\alpha =1$ and, by computing the cumulants explicitly, we shall verify in due course that this property
does indeed hold for our model.
\end{itemize}


\section{Beyond perturbations~: explicit form of the secular equation}
\label{sec:SecularEquation}

As explained in \S \ref{sec:GLE},  our approach is to reduce the spectral problem to the solution of the secular equation \eqref{eq:GeneralSecularEq}, where
 the coefficients $\Omega_+(\Lambda)$ and $\Omega_-(\Lambda)$ are to be found by computing $\phi_+(s;\Lambda)$ and $\phi_-(s;\Lambda)$ respectively.
For the particular L\'evy exponent \eqref{eq:LevyCauchy}, the equations \eqref{eq:ZeDifferentialEquation} 
reads
\begin{equation}
  \label{eq:SpectralPbCauchy}
  \left[ 
    - \deriv{^2}{s^2} + \frac{q}{s}\deriv{}{s} + E \pm \I\,c  + \frac{\I\,\Lambda}{s}
    \right]
    \phi_\pm(s;\Lambda)
    =0
    \hspace{0.25cm}\mbox{for }
    \pm s>0
    \:.
 \end{equation} 
We begin by considering the ``$+$'' case:
a simple substitution reduces it to the confluent hypergeometric equation or alternatively, to the Whittaker equation \cite{gragra}. 
The solution that decays at $+\infty$ is
\begin{align}
  &\phi_+(s;\Lambda) = a\, 
  \left( 2 \kstar \, s \right)^{q/2}\,
  W_{-\I\Lambda/(2\kstar),(q+1)/2}\left( 2 \kstar \,s \right)
  \\
  & 
  = a\, 
  \left( 2 \kstar \, s \right)^{q+1}\,\EXP{-\kstar \, s}\,
    \Psi\left( \frac{\I\Lambda}{2\kstar} + \frac{q}{2} +1 , q +2 ; 2\kstar\,s \right)
\nonumber
\end{align}
where $\Psi$ is the confluent hypergeometric (Kummer) function \cite{gragra}.
Assuming $q > -1$, we select $a$ so that $\phi_+(0;\Lambda)=1$.
For $q=0$, the identity $\Psi(1,2;z)=1/z$ leads to $ \phi_+(s;0)=\hat{f}(s) =\EXP{-\kstar \, s}$, so this is consistent with the result obtained in the previous paragraph.
 
To compute $\Omega_+(\Lambda)$, it is more convenient to rewrite $\phi_+$ in terms of the regular Kummer function  
$\Phi(a,c;z)=\sum_{n=0}^\infty z^n\,(a)_n/\big[(c)_n\,n!\big]$.
This is readily achieved by using Formula 9.210 of \cite{gragra}; the result is
\begin{align}
  \label{eq:SolutionForPositiveS}
  \phi_+(s;\Lambda) =  & \:\EXP{-\kstar \, s } \:
  \bigg[
    \Phi\left( \frac{\I\Lambda}{2\kstar} - \frac{q}{2} , -q ; 2\kstar\,s \right)
    \nonumber  \\
    &
    + 
        \left( 2 \kstar \, s \right)^{q+1}       
    \frac{\Gamma(-q-1)\,\Gamma\Big(\frac{\I\Lambda}{2\kstar} + \frac{q}{2} +1 \Big)}
           {\Gamma(q+1)\,\Gamma\Big(\frac{\I\Lambda}{2\kstar} - \frac{q}{2}\Big)}
    \nonumber  \\
    &\hspace{0.5cm}\times       
    \Phi\left( \frac{\I\Lambda}{2\kstar} + \frac{q}{2} +1 , q +2 ; 2\kstar\,s \right)
  \bigg]
  \:.
\end{align}
This expression provides a concrete example of the general expansion \eqref{eq:omegaPlus}.
Upon inserting the MacLaurin expansion for  $\Phi$, we immediately deduce 
\begin{equation}
  \label{eq:OmegaOfLambda}
  \Omega_+(\Lambda) =  
  \left( 2 \kstar \right)^{q+1} 
           \frac{\Gamma(-q-1)\,\Gamma\Big(\frac{\I\Lambda}{2\kstar} + \frac{q}{2} +1 \Big)}
                {\Gamma(q+1)\,\Gamma\Big(\frac{\I\Lambda}{2\kstar} - \frac{q}{2}\Big)}
\end{equation}

As it is clear from Eq. \eqref{eq:SpectralPbCauchy}, the equation for $\phi_-$ can be deduced from the one for $\phi_+$ by changing the signs of $s$, $c$ and~$\Lambda$. Hence 
\begin{equation}
  \Omega_+(\Lambda) \ 
  \overset{\stackrel{\Lambda\to-\Lambda}{c\to-c}}{\longrightarrow}  \ 
  \Omega_-(\Lambda) 
\end{equation}
and this leads to the following explicit form of the secular equation \eqref{eq:GeneralSecularEq}:
\begin{equation}
\label{eq:SecularEqCauchy}
     \kstar ^{q+1} 
  \frac{\Gamma\Big(\frac{\I\Lambda}{2\kstar} + \frac{q}{2} +1 \Big)}
       {\Gamma\Big(\frac{\I\Lambda}{2\kstar} - \frac{q}{2}\Big)}
  =
  \left( \kstar^* \right)^{q+1}
  \frac{\Gamma\Big(\frac{-\I\Lambda}{2\kstar^*} + \frac{q}{2} +1 \Big)}
       {\Gamma\Big(\frac{-\I\Lambda}{2\kstar^*} - \frac{q}{2}\Big)}
\:.
\end{equation}
The fundamental result of this paper is that the roots of this transcendental equation yield the eigenvalues $\Lambda_n (q)$ of the operator $\mathscr{L}_q$. 


\subsection{A symmetry property of the GLE}
\label{subsec:SymmetryGLE}

The secular equation \eqref{eq:SecularEqCauchy} exhibits an obvious  symmetry~:
it is invariant under the transformation 
\begin{equation}
   q+1\to-q^*-1
   \hspace{0.5cm}\mbox{and} \hspace{0.5cm}
   \Lambda\to\Lambda^*
   \:.
\end{equation}
This property implies a symmetry of the full spectrum of eigenvalues
$\big\{\Lambda_n(q)\big\}_{n\in\mathbb{Z}}=\big\{\Lambda_n(-q^*-2)^*\big\}_{n\in\mathbb{Z}}$.
In fact, for a very natural ordering of the eigenvalues, the analysis to follow in \S~\ref{sec:FullSpectrum} suggests the more precise relationship
\begin{equation}
  \label{eq:GeneralizedVanneste}
  \Lambda_n(q) = \big(\Lambda_{-n}(-q^*-2)\big)^*
  \:.
\end{equation}
and we expect this relationship to hold for more general models \cite{ComTexTou19,Tex20}.
For $n=0$ and $q$ real, it reduces to
\begin{equation}
  \label{eq:SymmetryVanneste}
  \Lambda(q) = \Lambda(-q-2)
  \:.
\end{equation}
In Ref.~\cite{Van10}, it was argued that the symmetry property $\Lambda(q) = \Lambda(-q-2m)$ should hold quite generally for products of random $2m\times2m$ symplectic matrices.
Its occurence here comes from the fact that, as explained in the introduction, our model is formulated as a continuum limit of a product of matrices in the group $\mathrm{SL}(2,\mathbb{R})$, which coincides with the $2 \times 2$ symplectic group $\mathrm{Sp}(2,\mathbb{R})$. 
Some counter-examples however exist \cite{StuThi19,Tex20} and the precise conditions under which the relation \eqref{eq:SymmetryVanneste} is true remain to be clarified.


\section{Exact expressions for the first four cumulants}
\label{sec:FFC}

It is clear from Eqs.~(\ref{eq:DefGLE},\ref{eq:RightEV}) that the cumulants $\gamma_n$ can in principle be found by treating $q$ as a perturbation parameter. 
In the traditional approach, illustrated in \S~\ref{sec:PerturbationInQ}, one needs to keep track of the corresponding eigenfunction, and this usually results in formulae involving (multiple) integrals~\cite{SchTit02,Mon21}. 
In this section, we apply the perturbative approach to the secular equation itself; the problem of computing the eigenvalues and that of computing the eigenfunctions are decoupled, and this produces formulae for the cumulants that are free of integrals.

We can simplify the analysis by using the fact that $\Lambda(q)$ is real for $q$ real, so that the secular equation \eqref{eq:SecularEqCauchy} can be expressed as $\im[\Omega_+(\Lambda)]=0$, a form used in Refs.~\cite{Tex20,Tex20b}. 
Here it yields
\begin{equation}
\label{eq:SecularForExpansion}
      \im\left[
         \left( 
            \frac{\I\Lambda(q)}{q} - \kstar
         \right)\,         
         \kstar^{q}\:
         \frac{\Gamma\Big(1 + \frac{q}{2} + \frac{\I\Lambda(q)}{2\kstar}\Big)}
              {\Gamma\Big(1 - \frac{q}{2} + \frac{\I\Lambda(q)}{2\kstar}\Big)}
      \right]  
      =0
      \:.
\end{equation} 
The idea is then to expand the left-hand side in powers of $q$. Equating the coefficient of the $q^n$ term to zero then provides an equation for $\gamma_{n+1}$ in terms of lower cumulants. 

Now, for $\xi$ independent of $q$, we can write
\begin{align}
   &      \frac{\Gamma\Big(1 + \frac{q}{2} + \xi \Big)}
              {\Gamma\Big(1 - \frac{q}{2} + \xi \Big)}
  =  1 + \psi(1)\, q 
  + \left(\psi'(1)\,q\,\xi + \frac{\psi(1)^2}{2}q^2 \right)
  \nonumber\\
  &+  \left(
    \frac{\psi''(1)}{2}\,q\,\xi^2 +\psi(1)\,\psi'(1)\,q^2\xi
     +\left[\frac{\psi''(1)}{24}+\frac{\psi(1)^3}{6}\right]q^3 
    \right)
    \nonumber\\
  &+\mathcal{O}(q^4) 
\end{align}
where $\psi(z)$ here denotes the digamma function. 
We use $\psi(1)=-\mbox{\bf C}$, where $\mbox{\bf C} \simeq 0.577\ldots$ is the Euler-Mascheroni constant,
$\psi'(1)=\zeta(2)=\pi^2/6$ and $\psi''(1)=\zeta(3)$, where $\zeta(x)$ is the Riemann zeta function.
By setting $\xi=\I\Lambda(q)/\big(2\kstar\big)$ and expressing $\Lambda(q)$ in terms in the cumulants,
we obtain the desired expansion for the left-hand side of Equation \eqref{eq:SecularForExpansion} in powers of $q$.
As is typical of such calculations, the complexity increases rapidly with $n$. The results for the first four cumulants are summarised below.

\paragraph*{The Lyapunov exponent and the variance~:}
The $q^0$ term of the secular equation \eqref{eq:SecularForExpansion} is obviously
\begin{equation}
      \im\left[ \I \,\gamma_1 - \kstar \right] = 0  
      \:.
\end{equation}
Using the fact that $\I \,\gamma_1 - \kstar =-\pi\,\IDoS$ is real, we see that the $q^1$ term of Equation \eqref{eq:SecularForExpansion} yields
\begin{equation}
   \gamma_2 =  \pi \IDoS \,\im\left[ 2\ln \kstar \right]\,.
\end{equation}
Since $2\ln \kstar=2\ln|\kstar|+\I\,\thetastar$,
these formulae for $\gamma_1$ and $\gamma_2$ agree with our previous calculations, which used the perturbative approach of \cite{Tex20,ComTexTou19}; see Equations (\ref{eq:Gamma1Cauchy},\ref{eq:Gamma2Cauchy}).

\paragraph*{Third cumulant~:}
Calculation of the $q^2$ term leads to
\begin{equation}
\label{eq:Gamma3Cauchy}
  \gamma_3 
  = \frac{\pi^2|\kstar|}{2}\,
  \cos^2\left(\frac{\thetastar}{2}\right)\,
  \sin\left(\frac{\thetastar}{2}\right)
  =  \frac{\pi^2c}{4} \sqrt{\frac{\sqrt{E^2+c^2}+E}{2(E^2+c^2)}}
 \:.
\end{equation}
In particular, $\gamma_3\simeq(\pi^2/2)\,\gamma_1$ as $E\to+\infty$.

\paragraph*{Fourth cumulant~:}
For the $q^3$ term, some complicated algebra eventually leads to
\begin{align}
\label{eq:Gamma4Cauchy}
  \gamma_4
  = &|\kstar|\,\cos\left(\frac{\thetastar}{2}\right)
  \bigg[
    \pi^2 \thetastar\,\cos^2\left(\frac{\thetastar}{2}\right)
    \nonumber\\
    &+ \thetastar ^3
    - 6\zeta(3) \,\sin(\thetastar)\sin^2\left(\frac{\thetastar}{2}\right)
  \bigg]
  \:.
\end{align}
In particular, $\gamma_4\simeq2\pi^2 \,\gamma_1$ for $E\to+\infty$.

\begin{figure}[!ht]
\centering
\includegraphics[width=0.475\textwidth]{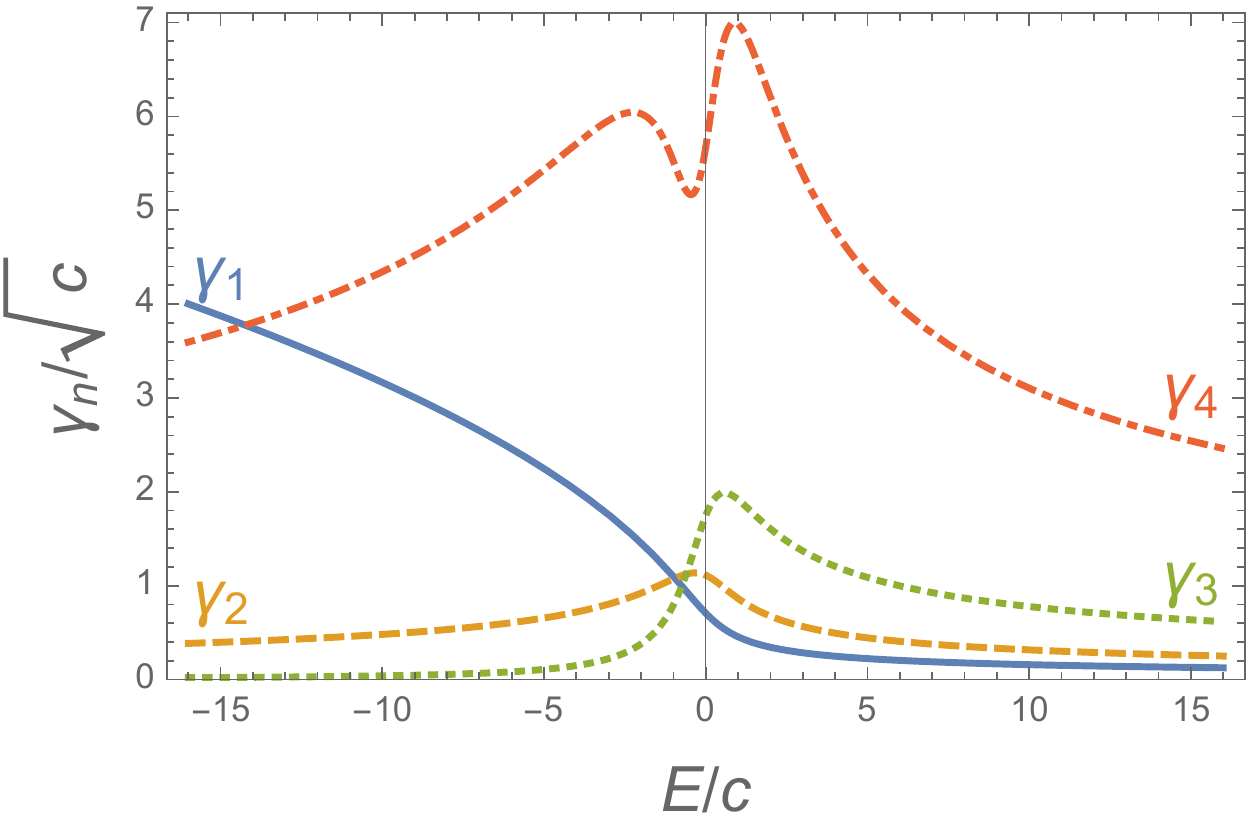}
\caption{\it   The four first cumulants as a function of the energy.}
\label{fig:CumulantsCauchy}
\end{figure}

The first four cumulants are plotted against the energy $E$ in Fig.~\ref{fig:CumulantsCauchy}.

We end by discussing briefly two limits~:
\begin{itemize}[label={$\bullet$},leftmargin=*,align=left,itemsep=0.1cm]
\item {\bf $E=+k^2\to+\infty$~:} For large positive energy, our calculations yield 
\begin{equation}
  \gamma_1\sim\gamma_2\sim \gamma_3\sim \gamma_4\sim \mathcal{O}(c/k)
\end{equation}
and so the fluctuations are \textit{non-Gaussian}.  
This is very different from the case of disorder with finite moments $\mean{v_n^2}<\infty$ (i.e. L\'evy exponent $\levy(s)\sim s^2$ for $s\to0$), for which we have $\gamma_n\ll\gamma_1\simeq\gamma_2$ for $n>2$ in the weak-disorder limit \cite{AntPasSly81,SchTit02,SchTit03,Tex20b}.

\item{\bf $E=-k^2\to-\infty$~:} For large negative energy,  we have instead
$\gamma_1\simeq\sqrt{|E|}$,
$\gamma_2\simeq (\pi/2)\,c |E|^{-1/2}$,
$\gamma_3\simeq (\pi^2/8)\,c^2|E|^{-3/2}$ and 
$\gamma_4\simeq (\pi^3/2)\,c |E|^{-1/2}$.  Thus
\begin{equation}
 \gamma_1-\sqrt{|E|}\sim\gamma_3 \sim \mathcal{O}(c^2/k^3)
 \ll  \gamma_2\sim \gamma_4  \sim \mathcal{O}(c/k)
  \:.
\end{equation} 
In this limit, the fluctuations are symmetric about the mean, but again non-Gaussian. 

\end{itemize}


\section{The universal ($E/c\to+\infty$) regime}
\label{sec:UniversalRegime}


\subsection{Explicit limiting form of the GLE}
\label{subsec:ELFofGLE}

In the high-energy regime, the previous analysis suggests that the cumulants all scale in the same way with the energy $\gamma_n\sim\gamma_1\simeq c/(2\sqrt{E})$. This motivates the introduction of the dimensionless rescaled GLE
\begin{equation}
   \lambda^+(q)\eqdef\lim_{E/c\to+\infty}\frac{\Lambda(q)}{\gamma_1}
\end{equation}
so that $\Lambda(q) \simeq \lambda^+(q)\,c/\big(2\sqrt{E}\big)$, and of
the dimensionless parameter $\epsilon=c/E$.
In terms of these new variables, the secular equation \eqref{eq:SecularForExpansion} takes the form
\begin{align}
  \label{eq:ConvenientSecularEq}
  \im\bigg[
      \left( \frac{\I\epsilon}{2}\frac{\lambda^+(q)}{q} - \sqrt{1+\I\epsilon} \right)
      &\left(1+\I\epsilon\right)^{q/2}
      \\
      \nonumber
      &\times
      \frac{\Gamma\Big(1+\frac{q}{2}+\frac{\I\epsilon \lambda^+(q)}{4\sqrt{1+\I\epsilon}}\Big)}
           {\Gamma\Big(1-\frac{q}{2}+\frac{\I\epsilon \lambda^+(q)}{4\sqrt{1+\I\epsilon}}\Big)}
  \bigg]=0\,.
\end{align}
In the previous section, we found expressions for the cumulants $\gamma_n$ by expanding this equation in powers of $q$, and we ascertained their
high-energy behaviour by taking the limit $\epsilon=c/E\to0$ in the resulting expressions. 
Here, in the spirit of \cite{Tex20b}, we reverse the order of the two limits: 
we first let $\epsilon\to0$ and obtain an expression for the GLE in the universal high-energy regime; in the next subsection, we use the expression obtained to deduce the asymptotics of the cumulants in this regime.

To proceed, we remark that Equation \eqref{eq:ConvenientSecularEq} can be expanded in powers of $\epsilon$ by using
$\Gamma(1\pm q/2+\xi)=\Gamma(1\pm q/2)\,\big[1+\xi\,\psi(1\pm q/2)+\mathcal{O}(\xi^2)\big]$. 
The $\epsilon^0$ term of the secular equation yields
\begin{equation}
  \lambda^+(q)
  = \frac{2(1+q)}{\psi\big(1-\frac{q}{2}\big)-\psi\big(\frac{q}{2}\big)}
  \:.
\end{equation}
This can be expressed in terms of elementary functions by using the identity
\begin{equation}
  \label{eq:FromComplementFormula}
  \psi(1-z) - \psi(z) = \pi \cotg \pi z
\end{equation}
which follows from the complement formula $\Gamma(z)\Gamma(1-z)=\pi/\sin(\pi z)$. The upshot is
\begin{equation}
  \label{eq:GLEUniversalLimit}
   \lambda^+(q)
   = \frac{2}{\pi}\,(1+q)\,\tan\left(\frac{\pi q}{2}\right)
   \:.
\end{equation}
In \cite{Tex20b}, this formula was obtained as a limiting case of a more general formula for power law disorder, itself derived by a perturbative method in the weak disorder limit.
Here we followed a different route and recovered this expression from our exact secular equation \eqref{eq:SecularEqCauchy} corresponding to Cauchy disorder.
We observe that the right-hand side blows up when $q$ is an odd integer. 
So the largest interval containing $0$ for which this formula makes sense is
\begin{equation}
  q\in]-3,1[
  \:.
\end{equation}
We see in particular that the GLE is positive for $q\in]-3,-2[\cup]0,1[$--- implying an exponential \textit{growth} of the moment $\smean{|\psi(x)|^q}$ with $x$ and negative for $q\in]-2,0[$, corresponding to an exponential \textit{decay} of the moment.

The most striking feature of our formula is obviously the fact that the GLE--- equivalently the moment $\mean{|\psi(x)|^q}$--- blows up at the endpoints of the finite interval $]-3,1[$; see
Fig.~\ref{fig:NumLambda1}.
This behaviour, characteristic of power-law disorder, is in sharp contrast with that observed in the case of the Halperin model, where the potential is a Gaussian white noise
corresponding to the L\'{e}vy exponent $\levy(s)=(\sigma/2)\,s^2$. In the Halperin model, the GLE exists for every $q$ and it was shown in \cite{FyoLeDRosTex18,Tex20} 
that it behaves asymptotically like $\Lambda^\mathrm{(Halp)}(q)\simeq(3/4)\,(\sigma/2)^{1/3}\,|q|^{4/3}$ as $q\to\pm\infty$. This behaviour was identified earlier by Bouchaud {\em et al.} \cite{BouGeoLeD86}, who made use of the replica trick. 

The simple analytic formula \eqref{eq:GLEUniversalLimit} is another important result of this paper. 
It will enable us to derive a general formula for the cumulants, in the high-energy limit, and to deduce the large deviation function controlling the distribution of the wave function. Furthermore, the formula provides a limiting form of the GLE
that is expected to be \textit{universally} valid in the weak-disorder/high-energy limit for models in which the disorder is characterised by the power-law tail $p(V)\sim V^{-2}$ for $V\to\pm\infty$ 
\cite{footnote7}.
For instance, we shall see in the next paragraph that it correctly reproduces the few results concerning the cumulants that are known for the Lloyd lattice model.


\subsection{Cumulants in the universal regime}

The rescaled GLE
\begin{equation}
\label{eq:universalCumulantGeneratingFunction}
   \lambda^+(q) 
  = \sum_{n=1}^\infty\frac{\CumPos_n}{n!}q^n 
\end{equation}
is the generating function of the rescaled cumulants $\CumPos_n$; they are related to the $\gamma_n$ via
\begin{equation}
  \gamma_n \simeq \CumPos_n\,\gamma_1 \simeq \frac{c}{2\sqrt{E}}\,\CumPos_n
\end{equation}
for $E/c\to+\infty$.
The MacLaurin expansion of the tangent function involves the Bernoulli numbers \cite{gragra}~:
\begin{equation}
  \tan x = \sum_{k=1}^\infty \frac{2^{2k}(2^{2k}-1)|B_{2k}|}{(2k)!}x^{2k-1} \:.
 \end{equation}
Thus
\begin{equation}
  \label{eq:CumulantsCauchyCase}
   \CumPos_n =    
   \begin{cases}
   4 \pi^{n-2}\left( 2^n -1 \right) |B_n|
   & \text{for $n$ even} \\
   \displaystyle\frac{\CumPos_{n+1}}{n+1}
   & \text{for $n$ odd} 
   \end{cases}
\end{equation} 
($\CumPos_1=1$ by definition).
Using $B_2=1/6$, $B_4=-1/30$, $B_6=1/42$, etc, we recover the relation $\CumPos_2 =2\simeq\gamma_2/\gamma_1$, 
i.e. Eq.~\eqref{eq:SPSCauchyWeak}, first obtained in Refs.~\cite{DeyLisAlt00,DeyLisAlt01}.
For the third, fourth and fifth cumulants, Titov \& Schomerus~\cite{TitSch03} found the numerical estimates
$\CumPos_3\simeq5$, $\CumPos_4\simeq20$ and $\CumPos_5\simeq100$.
Eq.~\eqref{eq:CumulantsCauchyCase} gives $\CumPos_3 = \pi^2/2 \simeq4.93$, $\CumPos_4=2\pi^2\simeq19.7$ and $\CumPos_5=\pi^4\simeq97.4$ in agreement with these estimates.

The large-$n$ behaviour of the cumulants can also be obtained by using the formula [24.11.1] of \cite{DLMF} 
$B_{2n}\sim(-1)^{n+1}2(2n)!\,(2\pi)^{-2n}$, leading to the asymptotics 
\begin{equation}
  \CumPos_n \underset{n\to\infty}{\simeq}
  \frac{8\,n!}{\pi^2}
  \:. 
\end{equation}
The series \eqref{eq:universalCumulantGeneratingFunction} has a unit radius of convergence. This is related to the fact that the distribution of $\Upsilon(x)=\ln|\psi(x)|$ has the exponential tail $\EXP{-\Upsilon}$ as $\Upsilon\to+\infty$ for Cauchy disorder.



\section{Beyond the universal regime}
\label{sec:BeyondUniversal}

\subsection{The limit $E\to0$}
\label{sec:ZeroEnergy}

\begin{figure}[!ht]
\centering
\includegraphics[scale=0.5]{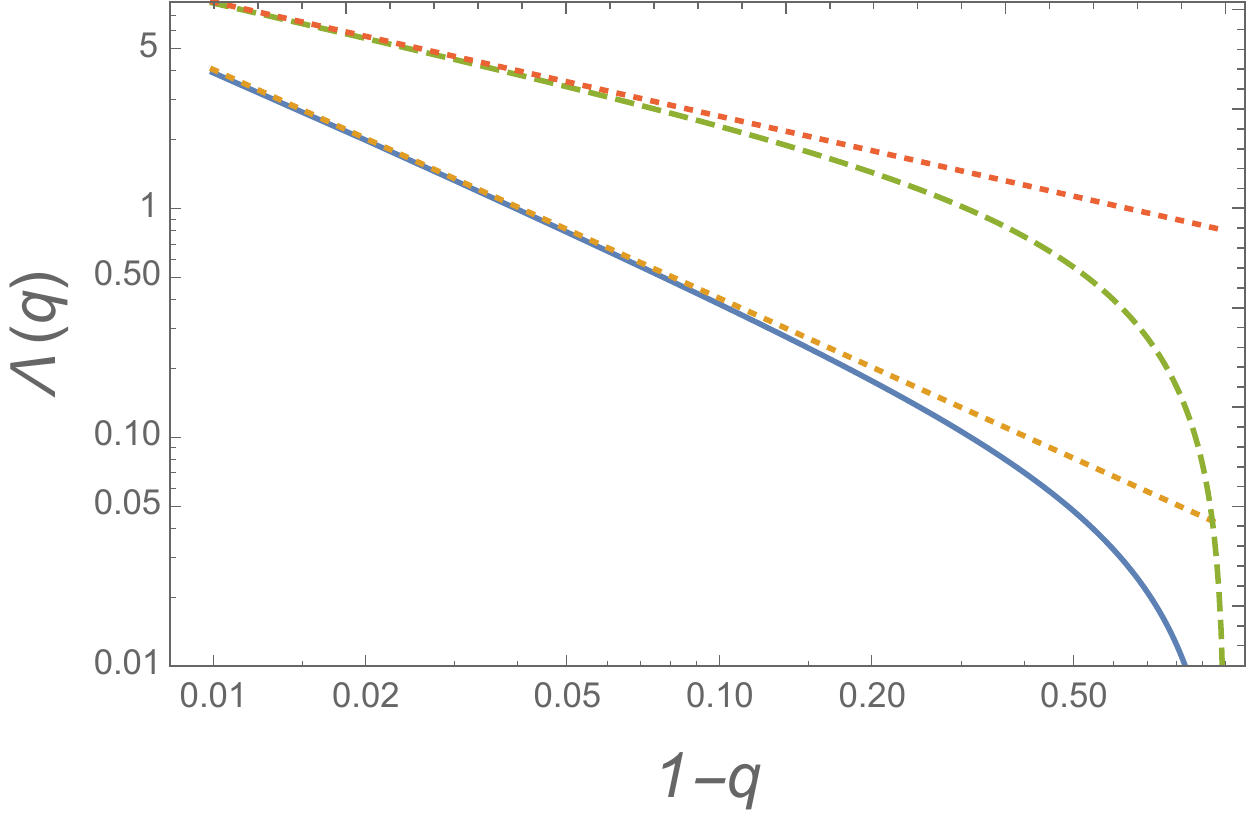}
\caption{\it Divergence of the GLE as $q\to1^-$,
for $k=10$ (blue continuous line) and $k=0$ (green dashed line).
The dotted lines are fits with  $\Lambda(q)\simeq 4c/\big[\pi^2k(1-q)\big]$ (orange)
and $\Lambda(q)\simeq0.8\sqrt{c/(1-q)}$ (red).
}
\label{fig:DivergenceGLE}
\end{figure}

The study of this limit is more involved; here, we content ourselves with a brief discussion of some results obtained by solving the secular equation~\eqref{eq:SecularForExpansion} numerically. 
These results suggest that the nature of the singularity at $q\to1^-$ (and thus, due to the symmetry \eqref{eq:SymmetryVanneste}, also as $q\to3^+$) depends on whether $E$ is large or small.
For $E=k^2 \gg c$, one verifies that the behaviour $\Lambda(q)\simeq 4c/\big[\pi^2k(1-q)\big]$ as $q\to1^-$---
which follows easily from Equation (\ref{eq:GLEUniversalLimit})--- is confirmed by the numerical analysis (see Fig.~\ref{fig:DivergenceGLE}).
The nature of the singularity changes when $k=0$~: the numerics (Fig.~\ref{fig:DivergenceGLE}) suggest the behaviour 
\begin{equation}
  \Lambda(q) \underset{q\to1^-}{\sim} \sqrt{\frac{c}{1-q}}
  \hspace{0.5cm}\mbox{for } k=0
  \:,
\end{equation}
and an analogous behaviour in the limit $q\to-3^+$.
We do not yet have an analytical proof for this limiting behaviour.

\subsection{The $E/c\to-\infty$ regime}
\label{sec:NegativeEnergy}

The limit of large negative energy is also of interest. The fact that, 
in this limit, $\gamma_4\simeq\pi^2\gamma_2\gg\gamma_1-\sqrt{|E|}\sim\gamma_3$,
suggests the rescaling 
\begin{equation}
  \lambda^-(q) \eqdef
  \lim_{E/c\to-\infty}\frac{\Lambda(q)-q\gamma_1}{\gamma_2}
  = \lim_{E/c\to-\infty} 
  \frac{\Lambda(q) - q\sqrt{|E|}}{  {\pi c}/\big({2\sqrt{|E|}}\big)  }
\end{equation}
Using $\kstar=\I\sqrt{|E|}\sqrt{1-\I\epsilon}$, where $\epsilon=c/|E|\ll1$, some simple algebra shows that the secular equation \eqref{eq:SecularForExpansion} assumes the form
\begin{align}
  \re\bigg[
    \I^q
    &\left(
      1 - \sqrt{1-\I\epsilon} + \epsilon\frac{\pi \lambda^-(q)}{2q}
    \right)
    (1-\I\epsilon)^{\frac{q}{2}}
    \nonumber\\
    &
    \times
    \frac{\Gamma\Big(1+\frac{q}{2} + \frac{q+(\epsilon\pi/2) \lambda^-(q)}{2\sqrt{1-\I\epsilon}}\Big)}
         {\Gamma\Big(1-\frac{q}{2} + \frac{q+(\epsilon\pi/2) \lambda^-(q)}{2\sqrt{1-\I\epsilon}}\Big)}
  \bigg]=0\,.
\end{align}
In the limit of small $\epsilon$, this yields
\begin{equation}
  \label{eq:Lminus}
  \lambda^-(q) = \frac{q}{\pi}\,\tan\left(\frac{\pi q}{2}\right)
   \hspace{0.5cm}\mbox{for }
  q\in]-1,1[
  \:.
\end{equation}

The interval of validity of this formula prompts us to make the following remark: expressed in terms of the rescaled GLE, the symmetry relation \eqref{eq:SymmetryVanneste} reads
\begin{equation}
  \lambda^-(-q-2) \simeq \lambda^-(q) + \frac{4|E|}{\pi c}\,(q+1)\,.
\end{equation}
Now, Equation \eqref{eq:Lminus} says that $\lambda_-(q)$ is finite for $q \in ]-1,+1[$. 
It follows from the symmetry relation that, for $q \in ]-3,-1[$, $\lambda_-(q)$ cannot have a limit as $|E|/c\to\infty$.

To end our discussion of this regime, we note that the rescaled cumulants $\CumNeg_n$ in the expansion
\begin{equation}
  \lambda^-(q) 
  = \sum_{n=1}^\infty\frac{\CumNeg_n}{n!}\,q^n
\end{equation}
are easily computed.
Indeed, the rescaled cumulants are related to the original ones by $\gamma_n\simeq\gamma_2\,\CumNeg_n$ for $E/c\to-\infty$.
Using, as before, the MacLaurin expansion of the tangent function, we deduce  
\begin{equation}
  \CumNeg_n = 
  \begin{cases}
   2 \pi^{n-2}\left( 2^n -1 \right) |B_n| & \text{for $n$ even} \\
   0 & \text{for $n$ odd}
   \end{cases}  \,. 
\end{equation}
In particular, we see that $\CumNeg_2=1$ and $\CumNeg_4=\pi^2$,  as they should, given the results obtained at the end of \S~\ref{sec:FFC}.
We also note the identity
\begin{equation}
  \CumPos_n=2\CumNeg_n
  \hspace{0.5cm}\mbox{($n$ even)}
\end{equation}
 relating the cumulants' behaviour as $E\to+\infty$  to their behaviour as $E\to-\infty$.


\section{The full spectrum of eigenvalues and the spectral gap}
\label{sec:FullSpectrum}

The study of the full spectrum of the operator $\mathscr{L}_q$ is interesting as it controls the spectral representation of the ``propagator'' \eqref{eq:PwithExp}
\begin{equation}
  \mathcal{P}_x(z|z_0;q) = \sum_n \Phi^\R_n(z;q)\,\Phi^\mathrm{L}_n(z_0;q)\,\EXP{x\,\Lambda_n(q)}
  \:,
\end{equation}
where $\Phi^\mathrm{L}_n(z;q)$ is the left-eigenvector associated with the problem adjoint to \eqref{eq:RightEV}.
The most important feature in this respect is the existence of a spectral gap 
\begin{equation}
  \Delta(q) = \re[\Lambda_0(q)-\Lambda_1(q)] >0\,.
\end{equation}
This is a crucial requirement in our approach: the ``propagator'' then behaves like
\begin{equation}
 \mathcal{P}_x(z|z_0;q) \simeq 
 \EXP{x\,\Lambda_0(q)}
 \left[ \Phi^\R_0(z;q)\,\Phi^\mathrm{L}_0(z_0;q) + \mathcal{O}\big( \EXP{-x\,\Delta(q)} \big) \right]
 \end{equation}
in the limit $x \rightarrow \infty$,
and Equation \eqref{eq:SpectralMethodToGetTheGLE}, which equates the GLE with the leading eigenvalue,
is then justified.


\subsection{The spectrum for $q=0$}

For this simple case, we find that
the solution set of the secular equation \eqref{eq:SecularEqCauchy} consists of the complex numbers
\begin{equation}
\label{eq:LambdaNforZeroQ}
  \Lambda_{n}(0) = 
  \begin{cases}
    -2\I\kstar^*\,n & \mbox{for } n\geq0
    \\
    -2\I\kstar\,n & \mbox{for } n<0
  \end{cases}
\end{equation}
The fact that $\Lambda_{-n}(0)=\Lambda_n(0)^*$ shows that the eigenvectors $\Phi_{n}^R(z;0)$ and $\Phi_{-n}^R(z;0)$ form a complex conjugate pair. 
The only real eigenvalue is $\Lambda_0(0)=0$. 
The gap is 
\begin{equation}
  \Delta(0) 
  = 2 | k_c | \sin\left(\frac{\thetastar}{2}\right)
  \,.
\end{equation}
By definition, $\theta_c$ is the argument of $E+\I c$ and so the gap is strictly positive since $c>0$.


\subsection{The universal regime}

In this regime, we can look for solutions of the secular equation \eqref{eq:SecularEqCauchy} in powers of $\epsilon=c/(2E)$~:
\begin{equation}
  \label{eq:LambdaNexpansion}
 \Lambda_n = \Lambda_n^{(0)} +  \frac{c}{2k}\,\ell_n + \mathcal{O}(c^2/k^3)
\end{equation}
where $E=k^2$ and $n$ labels the different solutions, i.e. the different eigenvalues of $\mathscr{L}_q$.
%
%
%
%
At lowest order $\epsilon^0$ (i.e. setting $c=0$), Eq.~\eqref{eq:SecularEqCauchy} reduces to
\begin{equation}
    \frac{\Gamma\Big(1 + \frac{\I\Lambda}{2k} + \frac{q}{2} \Big)}
       {\Gamma\Big(\frac{\I\Lambda}{2k} - \frac{q}{2}\Big)}
  =
  \frac{\Gamma\Big(1-\frac{\I\Lambda}{2k} + \frac{q}{2} \Big)}
       {\Gamma\Big(-\frac{\I\Lambda}{2k} - \frac{q}{2}\Big)}\,.
\end{equation}
Hence
\begin{equation}
  \sin\pi\left( \frac{q}{2}-\frac{\I\Lambda}{2k}\right)  = \sin\pi\left( \frac{q}{2}+\frac{\I\Lambda}{2k}\right)
\end{equation}
and the solutions of this equation are
\begin{equation}
  \Lambda_n^{(0)} = -2\I n \, k
    \hspace{0.5cm}
    \mbox{with }
    n\in\mathbb{Z}
    \:.
\end{equation}
As discussed in Subsection~\ref{subsec:SPwithRiccati}, these are the eigenvalues of the operator $\mathscr{D}_K(q)=\partial_{z}(k^2+z^2)+q\,z$ appearing in Equation \eqref{eq:SpectralPb} (see  \cite{Tex20}, Appendix B, and \cite{ComTexTou19}).


We can refine our calculation by 
inserting the expansion \eqref{eq:LambdaNexpansion} in the secular equation~\eqref{eq:SecularEqCauchy}~:
\begin{widetext}
\begin{align}
  \left( 1 + \I\epsilon \right)^{\frac{q+1}{2}}
  \frac{\Gamma(1+\frac{q}{2}+n)
        \left(1+\I\epsilon\left[-\frac{n}{2}+\frac{\ell_n}{4}\right]\psi(1+\frac{q}{2}+n)+\cdots\right)
       }{
         \Gamma(-\frac{q}{2}+n)
        \left(1+\I\epsilon\left[-\frac{n}{2}+\frac{\ell_n}{4}\right]\psi(-\frac{q}{2}+n)+\cdots\right)
         }
         \nonumber\\
       =  
  \left( 1 - \I\epsilon \right)^{\frac{q+1}{2}}
  \frac{\Gamma(1+\frac{q}{2}-n)
        \left(1-\I\epsilon\left[\frac{n}{2}+\frac{\ell_n}{4}\right]\psi(1+\frac{q}{2}-n)+\cdots\right)
       }{
         \Gamma(-\frac{q}{2}-n)
        \left(1-\I\epsilon\left[\frac{n}{2}+\frac{\ell_n}{4}\right]\psi(-\frac{q}{2}-n)+\cdots\right)
         }\,.
\end{align}
If we retain only the terms of order $\epsilon^1$, we deduce the following equation for $\ell_n$:
\begin{align}
  \label{eq:80}
  q+1 + \frac{n}{2}
  \left[
    - \psi\left(1+\frac{q}{2}+n\right)
    + \psi\left(-\frac{q}{2}+n\right)
    + \psi\left(1+\frac{q}{2}-n\right)
    -\psi\left(-\frac{q}{2}-n\right) 
  \right]
  \nonumber \\
  =
  \ell_n \left[
    - \psi\left(1+\frac{q}{2}+n\right)
    + \psi\left(-\frac{q}{2}+n\right)
    - \psi\left(1+\frac{q}{2}-n\right)
    +\psi\left(-\frac{q}{2}-n\right) 
  \right]\,.
\end{align}
\end{widetext}
To simplify the resulting formula, consider first the case $n>0$. By using \eqref{eq:FromComplementFormula} 
and also 
\begin{equation}
  \psi(z+n)-\psi(z-n) = (2z-1)\sum_{r=1}^n\frac{1}{(z+r-1)(z-r)}
\end{equation}
we eventually find
\begin{equation}
  \label{eq:NoIdeaForALabel}
  \ell_n 
  =\frac{2}{\pi}(q+1)\,\tan(\pi q/2)
  \left[
   1 - n\sum_{r=1}^n\frac{1}{(\frac{q}{2}+r)(\frac{q}{2}+1-r)}
  \right]\,.
\end{equation}
For $n<0$, a similar simplification can be achieved; we omit the details.
The upshot is the following formula for the eigenvalues, with an error of order ${\mathcal O}(\epsilon^2)$:
\begin{align}
 \label{eq:LambdaN}
  &\Lambda_n(q) \simeq -2\I n \, k
  \\
  \nonumber
  &+  \frac{c}{\pi\,k}(q+1)\,\tan(\pi q/2)
  \left[
   1 - |n|\sum_{r=1}^{|n|}\frac{1}{(\frac{q}{2}+r)(\frac{q}{2}+1-r)}
  \right]
\end{align}
with $n\in\mathbb{Z}$.
As a check, we consider the limit $q\to0$~: the expression \eqref{eq:LambdaN} become  $\Lambda_n(0)\simeq-2\I n\,k-(c/k)|n|$, in agreement with \eqref{eq:LambdaNforZeroQ}. 
Eq.~\eqref{eq:LambdaN} obviously satisfies the symmetry \eqref{eq:GeneralizedVanneste}.
At the symmetry point $q=-1$, we get 
\begin{equation}
 \label{eq:LambdaNunitary}
  \Lambda_n(-1) \simeq -2\I n \, k
  -  \frac{2c}{\pi^2k} 
  \left[
   1 + |n|\sum_{r=1}^{|n|}\frac{1}{(r-1/2)^2}
  \right]
  \:.
\end{equation}


\begin{figure}[!ht]
\centering
\includegraphics[scale=0.65]{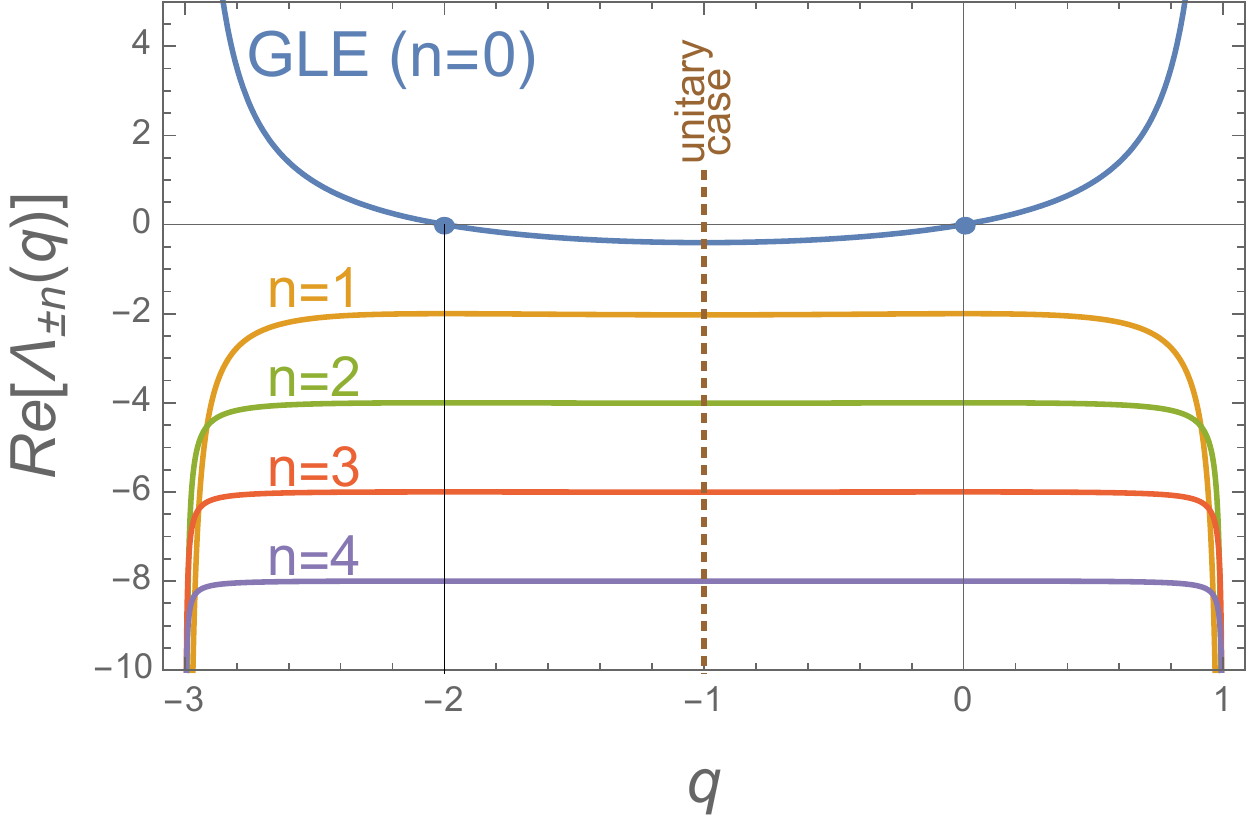}
\caption{
{\it The first five eigenvalues for $k=10$ (weak-disorder/high-energy universal regime).}
}
\label{fig:NumLambda1}
\end{figure}

In Fig.~\ref{fig:NumLambda1}, we have plotted the right-hand side of Equation \eqref{eq:LambdaN} against $q$ for $0\leq n \leq 4$. All the eigenvalues, except for the leading one, have a negative real part for every $q$. The gap is
\begin{equation}
  \Delta(q) 
  \simeq 
  \frac{c}{k\pi}\frac{4(q+1)\tan(\pi q/2)}{q(q+2)}>0
\end{equation}
and this is obviously positive in the interval $q \in ]-3,1[$.


\subsection{Unitary case ($q=-1$)}

As mentioned at the end of \S~\ref{subsec:SPwithRiccati}, the case $q=-1$ is of special significance because it corresponds to a case where the representation of $\mathrm{SL}(2,\mathbb{R})$ underlying the spectral problem can be made unitary.
It is also the symmetry point of the relation discussed in \S~\ref{subsec:SymmetryGLE}.

\begin{figure}[!ht]
\centering
\includegraphics[scale=0.5]{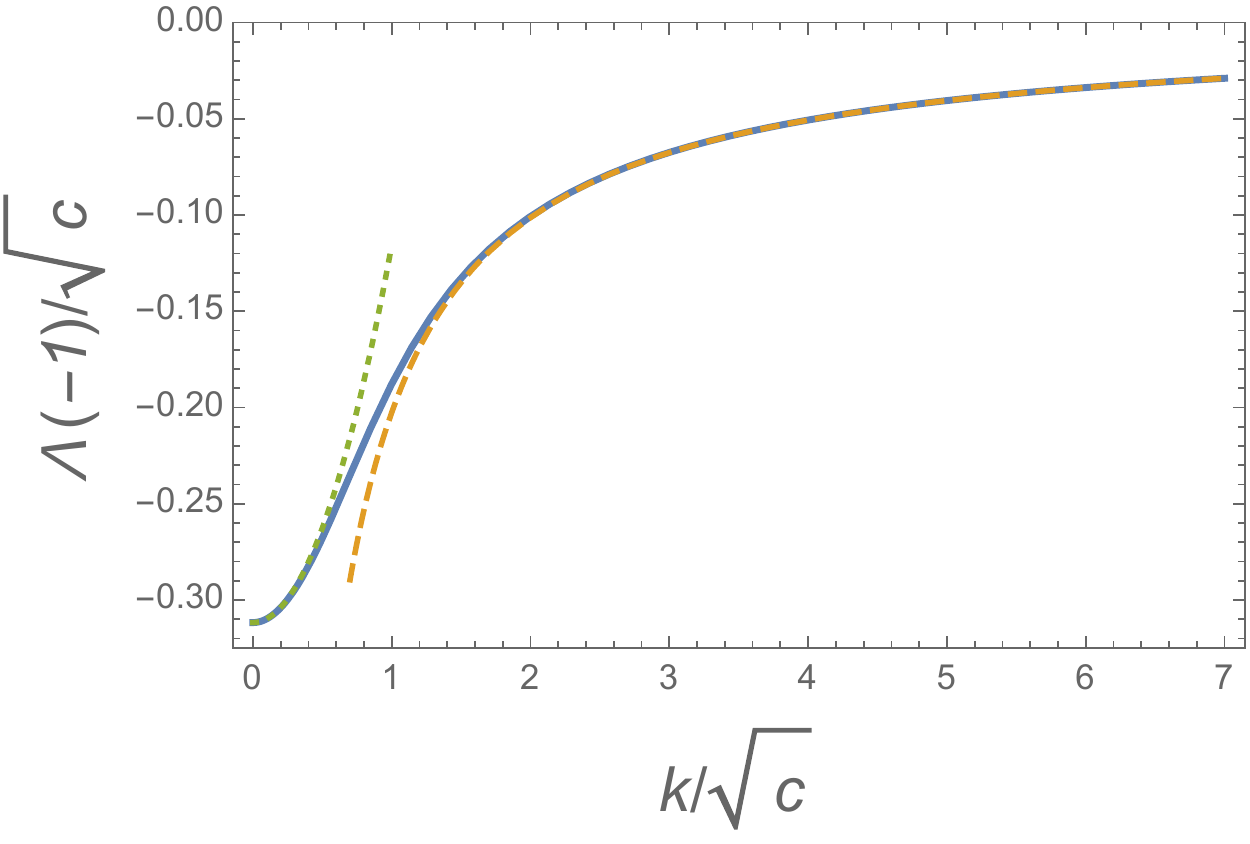}
\caption{
{\it $\Lambda(-1)$ as a function of $k=\sqrt{E}$. The orange dashed line is the perturbative result $\Lambda(-1)\simeq-2c/(\pi^2k)$ and the green dotted line $\Lambda(-1)\simeq a+b\,k^2$.}
}
\label{fig:LambdaUnitary}
\end{figure}

When we put $q=-1$ in the secular equation \eqref{eq:SecularEqCauchy}, it becomes a trivial identity.
In order to extract some information, we set $q=-1+\epsilon$ and expand the equation in powers of $\epsilon$.
At order $\epsilon^1$, we get 
\begin{align}
  \label{eq:SimplifiedSecularEq}
  \psi\left(\frac{1}{2} + \frac{\I\Lambda}{2\kstar} \right)
  -\psi\left(\frac{1}{2} - \frac{\I\Lambda}{2\kstar^*} \right)
  = -\I\,\arctan(c/k^2)
  \:.
\end{align}
where we have used $\ln \kstar-\ln \kstar^* = \I\,\thetastar$. 
The secular equation thus takes a simpler form in this case.
For $c\to0$, the ratio $\Lambda_0/\kstar\simeq\Lambda_0/k$ is almost real, so that we can use 
$\psi(1/2+\I x)-\psi(1/2-\I x) = \I\pi\tanh(\pi x)$. 
We get $\Lambda_0(-1)\simeq-2c/(\pi^2k)$ in agreeement with  the perturbative result, Eq.~\eqref{eq:LambdaNunitary}.
For $n=0$, the fact that $\Lambda\equiv\Lambda_0\in\mathbb{R}$ makes the equation \eqref{eq:SimplifiedSecularEq} easier to analyze.
In particular one readily finds the low-energy behaviour 
$\Lambda(-1)\simeq a+b\,E$ as $E\to0$ (the two coefficients can be determined numerically; the result is
$a\simeq-0.312\sqrt{c}$ and $b\simeq0.196/\sqrt{c}$).
The GLE is plotted in Fig.~\ref{fig:LambdaUnitary}.

\section{Wave function distribution and large deviation function}
\label{sec:LargeDev}

The GLE is the cumulant generating function for the logarithm of the wave function \cite{footnote8}
\begin{equation}
\Upsilon(x)=\ln|\psi(x)|
\:.
\end{equation}
We can therefore find its distribution by inverting a Laplace transform:  
\begin{align}
  P_x(\Upsilon) &\eqdef \mean{ \delta\left( \Upsilon - \ln|\psi(x)| \right) }
  = \int_{-\I\infty}^{+\I\infty} \frac{\D q}{2\I\pi} \,\EXP{-q\,\Upsilon}
  \,\mean{|\psi(x)|^q}
  \nonumber
  \\
  &\underset{x\to\infty}{\sim }
  \int_{-\I\infty}^{+\I\infty} \frac{\D q}{2\I\pi} \,\EXP{-q\,\Upsilon + x\, \Lambda(q)}\,.
\end{align}
This suggests the large deviation form 
\begin{equation}
  P_x(\Upsilon) \underset{x\to\infty}{\sim}
  \exp\left\{ -  x\: \LDF\!\left(\Upsilon/x \right) \right\}
\end{equation}
with
\begin{equation}
  \LDF\left( \xi \right)
  = \underset{q}{\mathrm{min}}\left\{ q\,\xi - \Lambda(q) \right\}\,.
\label{eq:legendre}
\end{equation}
The symmetry relation~\eqref{eq:SymmetryVanneste} implies \cite{Tex20}
\begin{equation}
\label{eq:LDFsymmetry}
  \LDF( -\xi ) = \LDF( \xi ) +2\xi
  \:.
\end{equation}

Note that the singularity of the GLE as $q\to1^-$ corresponds to an exponential tail $P_x(\Upsilon)\sim\EXP{-\Upsilon}$.
Indeed, assume a singularity of the form 
\begin{equation}
  \Lambda(q) \simeq 
  \frac{A}{\eta-1}\,(q_0-q)^{1-\eta}
    \hspace{0.5cm}\mbox{for }q\to q_0^-
    \:,
\end{equation}
with $\eta>1$.
Then the minimum in Equation \eqref{eq:legendre} is attained at $q_\ast \simeq q_0-(A/\xi)^{1/\eta}$, and so
  \begin{equation}
    \LDF(\xi) \simeq 
    q_0\xi - 
    \frac{\eta\,A^{1/\eta}}{\eta-1} \xi^{1-1/\eta}
    \hspace{0.5cm}\mbox{for }\xi\to+\infty
    \:.
  \end{equation}


\subsection{Universal weak disorder regime ($E/c\to+\infty$)}

\begin{figure}[!ht]
\centering
\includegraphics[scale=0.6]{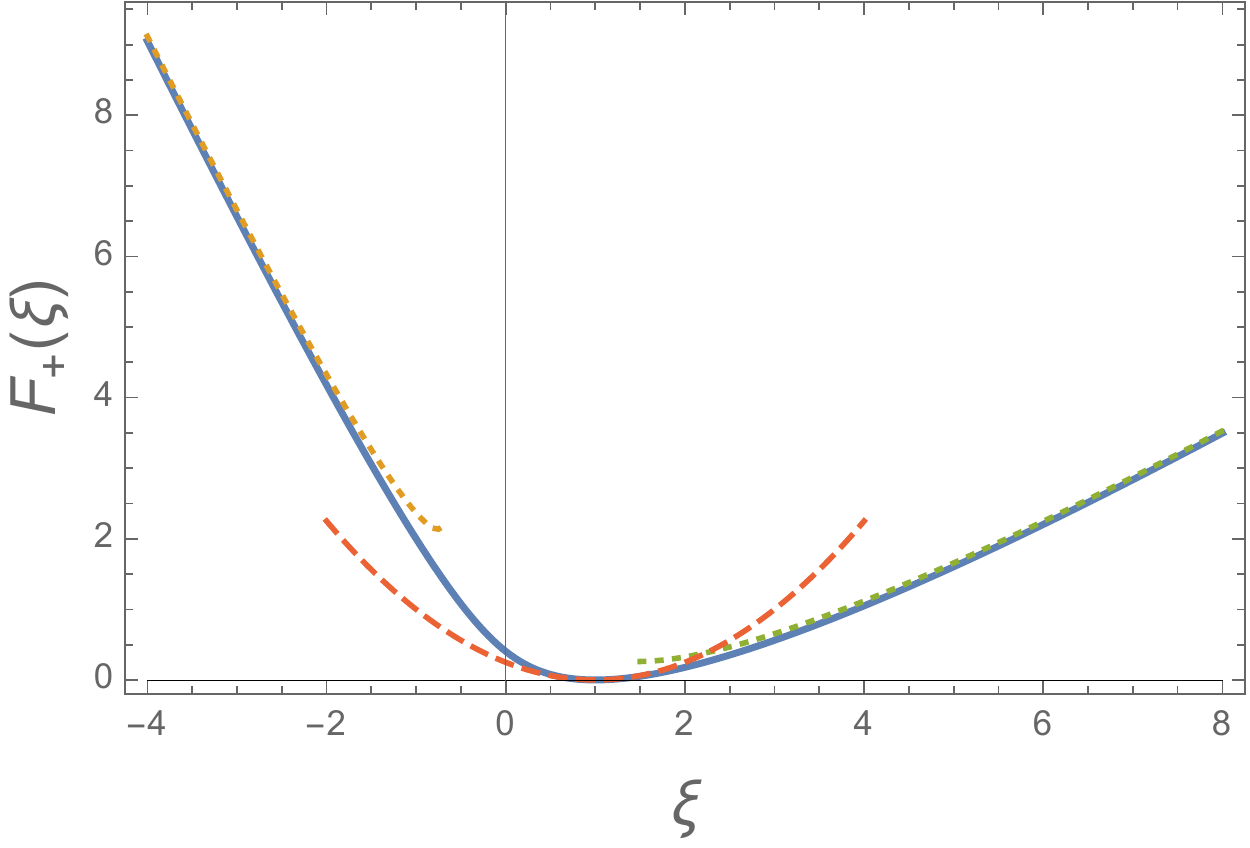}
\caption{\it The large deviation function controlling the distribution of $\ln|\psi(x)|$ in the regime $E\to+\infty$.
The dotted lines correspond to the limiting behaviours discussed in the text. }
\label{fig:Fplus}
\end{figure}

Using the result $\Lambda(q)\simeq\gamma_1\,\lambda^+(q)$, we expect the large deviation form
\begin{equation}
  \label{eq:DistribUpsilonUniversal}
  P_x(\Upsilon) \underset{x\to\infty}{\sim}
  \exp\left\{ - \gamma_1x\: \LDF_+\!\left(\frac{\Upsilon}{\gamma_1x} \right) \right\}
\end{equation}
where
\begin{equation}
  \LDF_+\left( \xi \right)
  = \underset{q}{\mathrm{min}}\left\{ q\,\xi - \lambda^+(q) \right\}
\end{equation}
is the dimensionless large deviation function.
The minimum is attained at $q_*$, the solution of   
\begin{equation}
  \xi = \frac{2}{\pi} \tan(\pi q_*/2) + \frac{q_*+1}{\cos^2(\pi q_*/2)}\,.
\end{equation}
Hence
\begin{align}
  \LDF_+\left( \xi \right) 
  &= \left( \frac{q_*+1}{\cos(\pi q_*/2)}\right)^2 - \xi
  \\
  &= \frac{q_*(q_*+1)}{\cos^2(\pi q_*/2)} - \frac{2}{\pi} \tan(\pi q_*/2)\,.
\end{align}
The function $\LDF_+(\xi) $ is plotted in Fig.~\ref{fig:Fplus}.

\paragraph{Typical values.---}

In the limit  $q_*\to0$, we can write
\begin{equation}
  \xi  = 1 + 2q_* +  \frac{\pi^2}{4} q_*^2 + \mathcal{O}(q_*^3)
\end{equation}
and 
\begin{equation}
   \LDF_+( \xi ) = 1+ 2q_*+q_*^2 +  \frac{\pi^2}{4} q_*^2 +\cdots - \xi
   = q_*^2 + \cdots
  \simeq \frac{1}{4}\left(\xi-1\right)^2\,.
\end{equation}
We recover the expected result for the variance, namely $\CumPos_2=2$.

\paragraph{Large deviations.---} 
 
 Setting $q=1-\epsilon$, some algebra gives the small-$\epsilon$ asymptotics
\begin{equation}
  \xi = \frac{8}{\pi^2\epsilon^2} + \frac23 +\mathcal{O}(\epsilon) 
\end{equation}
from which one deduces
$\epsilon\simeq\frac{4}{\pi}/\sqrt{2(\xi-2/3)}$. 
Thus
\begin{align}
  \LDF_+( \xi )  &\underset{\xi\to+\infty}{\simeq}  
  \frac{8}{\pi^2\epsilon^2} - \frac{16}{\pi^2\epsilon}  + \frac23 + \frac{4}{\pi^2} +\mathcal{O}(\epsilon) 
  \\
  &\simeq 
  \xi - \frac{4}{\pi}\sqrt{2(\xi-2/3)} +  \frac{4}{\pi^2}\,.
\end{align}
The tail associated with the limit $\xi\to-\infty$ can be analysed in a similar way by setting $q=-3+\epsilon$.

\paragraph{Summary.---}

The three limiting behaviours are 
\begin{equation}
  \label{eq:PhiPlusLimitingBehaviours}
  \LDF_+( \xi ) 
  \simeq
  \begin{cases}
     -3\xi - \frac{4}{\pi}\sqrt{2(-\xi-2/3)} +  \frac{4}{\pi^2} & \mbox{for } \xi\to-\infty
     \\[0.125cm]
         \frac{1}{4}\left(\xi-1\right)^2 & \mbox{for } \xi\sim1
     \\[0.125cm]
     \xi - \frac{4}{\pi}\sqrt{2(\xi-2/3)} +  \frac{4}{\pi^2} & \mbox{for } \xi\to+\infty
  \end{cases}\,.
\end{equation}
These limiting behaviours are consistent with the symmetry \eqref{eq:LDFsymmetry}.

The two behaviours  
$P_x(\Upsilon)\sim \EXP{+3\Upsilon}$ as $\Upsilon\to-\infty$ and 
$P_x(\Upsilon)\sim \EXP{-\Upsilon}$ as $\Upsilon\to+\infty$ are quite different from the ones obtained in Ref.~ \cite{FyoLeDRosTex18} for the Halperin model with a Gaussian white noise potential. 
In this case the distribution is 
$P_x(\Upsilon)\sim \exp\{-x\,\LDF^\mathrm{(Halp)}(\Upsilon/x)\}$ 
 where $\LDF^\mathrm{(Halp)}(\xi)$ is the Legendre transform of the GLE of the model; with the limiting behaviour recalled at the end of \S~\ref{subsec:ELFofGLE} for $\Lambda^\mathrm{(Halp)}(q)$, one gets
$\LDF^\mathrm{(Halp)}(\xi)\simeq\xi^4/(8\sigma)$ as $\xi\to\pm\infty$.
That is: $P_x^{(\mathrm{Halp})}(\Upsilon)\sim\exp\big\{-\Upsilon^4/(8\sigma x^3)\big\}$ as $\Upsilon\to\pm\infty$.

\subsection{Application: distribution of the conductance in the universal regime}
\label{sec:Conductance}

We deduce from \eqref{eq:PhiPlusLimitingBehaviours} the following limiting behaviours for the distribution of modulus of the wave function 
$\mathscr{P}_x(\psi) = \mean{\delta( \psi - |\psi(x)| )}=\frac1\psi\,P_x(\ln\psi)$~:
\begin{equation}
  \label{eq:WaveFctDistrib}
  \mathscr{P}_x(\psi) 
    \sim 
  \begin{cases}
     \psi^2 & \mbox{as } \psi\to0
     \\[0.125cm]
     \psi^{-1} \,\EXP{-\left( \ln \psi - \gamma_1 x \right)^2/(4\gamma_1 x)}
       & \mbox{as } \ln \psi \sim \gamma_1 x
     \\[0.125cm]
     \psi^{-2} & \mbox{as } \psi\to+\infty
  \end{cases}
\end{equation}
Let us elaborate the implication of this last formula for the distribution of the conductance.
According to the Borland conjecture \cite{Bor63}, the probability $g$ of transmission through a disordered sample of length $L$--- the dimensionless conductance of the sample---
should be related to the solution $\psi(x)$ of the initial value problem via
$
g \sim |\psi(L)|^{-2}\,.
$
This only holds for configurations with small transmission probability $g\ll1$; it does not describe the atypical configurations where the transmission probability is large, $g\lesssim1$.
Equation \eqref{eq:WaveFctDistrib} implies the following for the distribution $\mathcal{W}_L(g)$ of the conductance:
\begin{equation}
  \label{eq:ConductanceDistrib}
  \mathcal{W}_L(g)     
  \sim  
    \begin{cases}
     g^{-1} \,\EXP{-\left( \ln g +2 \gamma_1 L \right)^2/(16\gamma_1 L)}
     &  \mbox{as } \ln g \sim -2 \gamma_1 L
     \\[0.125cm]
     g^{-1/2} & \mbox{as } g\to0
  \end{cases}
  \:.
\end{equation}
The large deviation tail as $g\to1^-$ requires a different analysis.
The power law behaviour as $g\to0$ was demonstrated numerically in \cite{TitSch03}, but the precise value of the exponent was not determined.
The behaviour \eqref{eq:ConductanceDistrib} is in agreement with the numerical results of Mendez-Bermudez \textit{et al}. \cite{MenMarGopVar16} who conjectured the behaviour $\mathcal{W}_L(g)\sim g^{-1+\alpha/2}$ for the power law disorder $p(v)\sim|v|^{-1-\alpha}$.
This behaviour was later demonstrated for arbitrary~$\alpha$ by an analytic calculation in Ref.~\cite{Tex20b}. The case of Cauchy disorder corresponds to taking $\alpha=1$.

Although, in all cases, the typical values of the conductance are exponentially small, namely $g\sim\EXP{-2\gamma_1L}$, the power law singularity of $\mathcal{W}_L(g) $ at $g=0$ is completely different from that expected for the more standard case of disorder with finite moments.
To give a concrete example, for the Halperin model, we can deduce from the large deviation function $\LDF^\mathrm{(Halp)}(\xi)$ stated earlier that the conductance distribution exhibits  \textit{suppression} as $g\to0$~:
$
  \mathcal{W}_L^\mathrm{(Halp)}(g)
  \sim
  ({1}/{g})\,\exp\big\{ -\left(\ln g\right)^4  / (128\sigma L^3) \big\}
$.
Although the precise behaviour will depend on the details of the model, we expect such a suppression to be generic for disorder with finite second moment.

\subsection{Large deviation function for $E/c\to-\infty$}

In the limit $E\to-\infty$, the GLE assumes the form $\Lambda(q)\simeq q\sqrt{|E|}+\gamma_2\,\lambda^-(q)$. Hence we expect the large deviation form
\begin{equation}
  P_x(\Upsilon) \underset{x\to\infty}{\sim}
  \exp\left\{ - \gamma_2x\: \LDF_-\!\left(\frac{\Upsilon-\sqrt{|E|}\, x}{\gamma_2x} \right) \right\}
\end{equation}
where 
\begin{equation}
  \LDF_-\left( \xi \right)
  = \underset{q}{\mathrm{min}}\left\{ q\,\xi - \lambda^-(q) \right\}
  \:.
\end{equation}

\begin{figure}[!ht]
\centering
\includegraphics[scale=0.65]{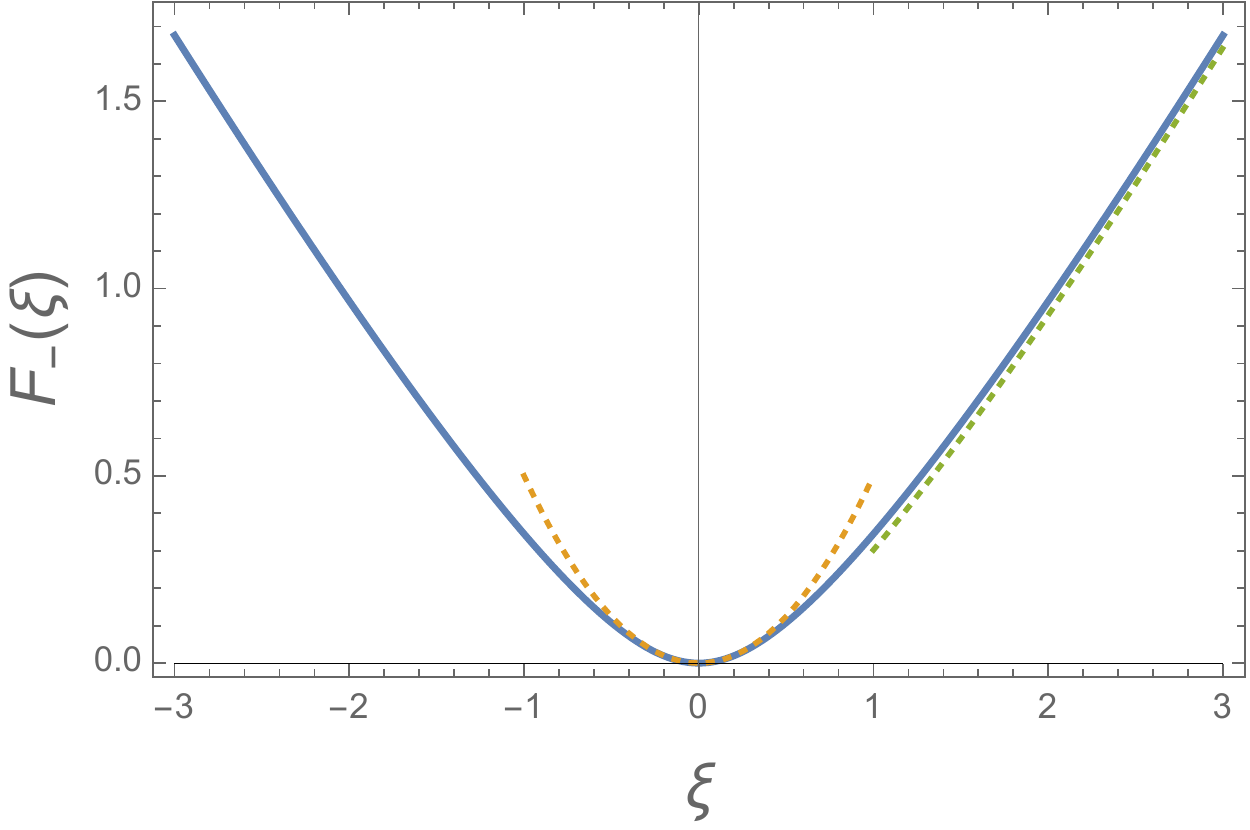}
\caption{\it The large deviation function controlling the distribution of $\ln|\psi(x)|$ in the regime $E\to-\infty$.
The dotted lines correspond to the asymptotics discussed in the text.
}
\label{fig:Fminus}
\end{figure}

Applying the Legendre transform yields 
\begin{equation}
  \xi = \frac{1}{\pi}\,\tan\left(\frac{\pi q_*}{2}\right)
  + \frac{q_*}{2\,\cos^2(\pi q_*/2)}
\end{equation}
and
\begin{equation}
  \LDF_-\left( \xi \right)
  = \frac{1}{2}\left( \frac{q_*}{\cos(\pi q_*/2)}\right)^2
  \:.
\end{equation}
Some algebra then gives the limiting behaviours
\begin{equation}
  \LDF_-\left( \xi \right) 
  \simeq
  \begin{cases}
     \frac12\xi^2  & \mbox{for } |\xi| \ll 1
     \\[0.125cm]
     |\xi| - \frac{2}{\pi}\sqrt{2|\xi|} + \frac{2}{\pi^2} +\mathcal{O}(\xi^{-1/2})  
                   & \mbox{for } |\xi| \gg 1
  \end{cases}\,.
\end{equation}
This translates into  
$P_x(\Upsilon)\sim \exp\big\{-|\Upsilon-\gamma_1x|\big\}$ for large~$\Upsilon$. 
The function $\LDF_-(\xi) $ is plotted in Fig.~\ref{fig:Fminus}.


\section{The spectral problem in terms of generalized Coulomb problems}
\label{sec:Alain}

The spectral problem \eqref{eq:SpectralPbFourier}-\eqref{eq:GeneralSecularEq} is unusual in many respects.
The purpose of this section is to explain its relationship with some recent 
work on certain generalizations of the spectral problem for the Schr\"{o}dinger equation with a Coulomb potential \cite{DerGeo20,DerFauNguRic20}. 

The calculation of the coefficients $\Omega_+$ and $\Omega_-$ that appear in the secular equation relied on the explicit solution of Equation \eqref{eq:SpectralPbCauchy}. For definiteness, consider the ``$+$'' case and set $\phi_+(s;\Lambda)= s^{q/2}\,\chi_+(s)$, so that the equation
for the new unknown $\chi_+$ is
\begin{equation}
  \label{eq:Alain1}
  L_{m,\beta}\,  \chi_+(s)= -k^2\chi_+ (s)
  \hspace{0.25cm}\mbox{for }
  s>0
\end{equation}
where 
\begin{equation}
  \label{eq:Alain2}
   L_{m,\beta} = -\deriv{^2}{s^2}
    + \left( m^2-\frac{1}{4} \right)\frac{1}{s^2}-\frac{\beta}{s}
    \:,
\end{equation}
and, in order to conform to the notation used in \cite{DerFauNguRic20}, we have introduced
$$
m = \frac{q+1}{2}\,,\;\;
\beta = - \I \Lambda\;\;\text{and}\;\;
k^2 = E + \I c\,.
$$
This has the same form as the differential equation for the radial part of the wave function describing a quantum particle in a Coulomb potential \cite{Eve05}. 
The classical example is the hydrogen atom, where the spectral problem of interest consists of finding the values of the energy $k^2$ such that $\chi_+$ is square-integrable.
However, the problem discussed in the present paper differs from the classical Coulomb problem in three respects~:
\begin{enumerate}[label={(\arabic*)},leftmargin=*,align=left,itemsep=0.1cm]
  \item   
  In the Coulomb problem, the point spectrum is the set $\{- k_n^2 \}$ such that $\chi^+(s)$ is square-integrable. By contrast, in our case $k^2=E+\I c$ is just a given parameter, and the spectrum consists of the numbers 
  $\Lambda_n$ for which the operator $L_{m,\beta}$, subject to some boundary conditions, has $-k^2$ as one of its eigenvalues.
Therefore, our problem  can be described as a {\em spectral problem in the coupling constant}.
 
  \item Some of the coefficients--- in particular the coupling constant $\beta=-\I\Lambda$--- are complex, so that the problem cannot be treated within the standard framework of self-adjoint operator theory.
 
  \item A third difference is the fact that our problem is defined on the whole real line.
\end{enumerate}
Nevertheless, it was shown recently in \cite{DerGeo20,DerFauNguRic20} that the familiar self-adjoint theory works almost as well in the complex case, provided the boundary
condition at $s=0$ is chosen from a certain family parametrized by the number $\kappa$.
In particular, for the so-called ``holomorphic'' family characterised by the triplet $\{\beta,\,m,\,\kappa\} $ with $\beta\,,\kappa \in\mathbb{C}$ and $-1<\re(m)<1$, the operator $L_{m,\beta}$,
supplemented with the boundary  condition 
\begin{align}
  \label{eq:Alain3}
  \chi^+(s) \sim \:
  &
    s^{1/2+m} \left( 1-\beta\frac{s}{1+2m} \right)
    \\ \nonumber 
  &+ \kappa \, s^{1/2-m} \left( 1-\beta\frac{s}{1-2m}\right)
  \mbox{ for }s\to0^+
\end{align}
is a closed operator in the space of square-integrable functions on the positive half-line. Furtheremore,
it is shown in \cite{DerFauNguRic20}  that the spectral problem for this operator is well-posed, with a complex point spectrum that coincides with the solution set of the
transcendental equation
\begin{equation}
  \label{eq:Alain4}
  \kappa= (2k)^{-2m}
  \frac{\Gamma(2m)\,\Gamma(1/2-m-\beta/(2k))}{\Gamma(-2m)\,\Gamma(1/2+m-\beta/(2k))}\,.
\end{equation}
In other words, for a fixed triplet $\{ \beta, \,m,\,\kappa\}$, the eigenvalues $k^2$ of the operator correspond to the values of $k$ that solve this equation.

The relevance of this theory to the calculation of the coefficient $\Omega_+$ becomes clear when, upon comparing Equations \eqref{eq:Alain3} and \eqref{eq:omegaPlus}, we realize that
$\kappa = 1/\Omega_+$. By requiring that $\phi_+$ decay at infinity, we are effectively imposing that
$E+\I c$ must be an eigenvalue of $L_{m,\beta}$, for the boundary condition \eqref{eq:Alain3}. So Formula \eqref{eq:OmegaOfLambda} could have been 
deduced from Formula \eqref{eq:Alain4}. Similar considerations apply to the coefficient $\Omega_-$.

We end by pointing out another interesting interpretation of the coefficient $\Omega_\pm$ in terms of the {\em Weyl--Titchmarsh coefficient} associated with a certain 
singular Sturm--Liouville operator on a half-line \cite{KurLug11}.
Recall that, for the operator 
$H= -\partial_x^2+V(x)$, with a potential that is well-behaved at $x=0$, we may express any solution of $H \psi = \lambda \psi$ as a linear combination
of two particular solutions, say
$f(x,\lambda)$ and $g(x, \lambda)$, satisfying
 \begin{equation}
   f(0,\lambda)=g'(0, \lambda)=0
   \hspace{0.25cm}\mbox{and}\hspace{0.25cm}
   g(0, \lambda)=-f'(0,\lambda)=1
   \:.
 \end{equation}
By definition, the Weyl--Titchmarsh coefficient is the unique number $w(\lambda)$ that makes the linear combination
\begin{equation}
  \label{eq:Alain6}
  \varphi(x,\lambda)= g(x,\lambda)-w(\lambda)\,f(x,\lambda)
\end{equation}
square-integrable.
When, as in the Coulomb case, the potential $V(x)$ is singular, a Weyl--Titchmarsh coefficient may still be defined, provided we adjust the conditions that $f(x,\lambda)$ and $g(x,\lambda)$ must satisfy at $x=0$ appropriately.  
The Weyl--Titchmarsh coefficient  $w(\lambda)$ computed in \cite{KurLug11} is essentially the coefficient $\kappa$ defined in Eq.~\eqref{eq:Alain4}. 
It is also clear that our analysis of the local behaviour of the solution of \eqref{eq:ZeDifferentialEquation}, valid for a generic L\'evy process, is also consistent
with the interpretation of $\Omega_\pm$ in such terms, so that the
secular equation \eqref{eq:GeneralSecularEq} can be understood as the equality between two Weyl--Titchmarsh coefficients: one associated with 
the ``$+$'' part of Equation \eqref{eq:SpectralPbCauchy}, and the other associated with the ``$-$'' part. This interpretation could shed some light on the spectral properties of the operator $\mathscr{L}_q$.


\section{Conclusion}
In this paper, we have addressed the question of wave function fluctuations for the one-dimensional Schr\"odinger equation with a random potential. We studied in great detail a case where the random potential is a L\'evy noise, and the underlying L\'evy process is the Cauchy process--- resulting in a model with power law disorder.
By building on recent progress in our understanding of the role of representation theory in the study of products of random matrices in $\text{SL}(2,{\mathbb R})$, we have been able to derive an explicit transcendental equation for the cumulant generating function--- also known as the generalized Lyapunov exponent (GLE)--- that describes the fluctuations.
This led to analytical expressions for the first few cumulants. We have also obtained the limiting form of the GLE and of all the cumulants in the high energy/weak disorder regime.
Considering the scarcity of exact results in this area, we view these findings as a significant progress.

Although much of the paper has been devoted to a particular model, our results have wider implications for the \og single parameter scaling \fg{} (SPS) conjecture.
In the more standard case where the moments of the disorder are finite, SPS manifests itself in the fact that $\gamma_1\simeq\gamma_2\gg\gamma_n$ for $n>2$, and so the large deviations (the atypical fluctuations) involve a scale that is different from that which controls the typical fluctuations.
For example, for the Schr\"odinger equation with a Gaussian white noise potential (the Halperin model), with $\psi$ the corresponding wave function, the mean value $\mean{\ln|\psi(x)|}\simeq\gamma_1 x$ and the typical fluctuations $\big(\ln|\psi(x)|\big)_\mathrm{typ.}\sim\sqrt{\gamma_1 x}$ are both controlled by the Lyapunov exponent, whilst the large deviation that characterizes atypically large fluctuations yields $\big(\ln|\psi(x)|\big)_\mathrm{atyp.}\sim (\sigma^{1/3}x)^{3/4}$.
This result is in sharp contrast with the case of Cauchy disorder studied in the present paper: we have shown that all the cumulants scale in the same way with energy and disorder strength in the high energy/weak disorder regime: $\gamma_n\sim c/\sqrt{E}$. As a consequence both typical and atypical fluctuations are controlled by the same scale.
This is a very strong manifestation of SPS,  characteristic of models with power law disorder \cite{Tex20b}. 
Indeed, by studying our continuous model, we have recovered the few known results from previous studies of the well-known tight-binding Lloyd model. This is a strong confirmation of the universal character of our results.

Our study has focused on the case of Cauchy disorder. It would be interesting to find other cases that could be solved via a secular equation~: for models where the noise arises from a L\'evy process, this requires finding a L\'evy exponent $\levy(s)$ such that the Weyl--Titchmarsh coefficients associated with each of the differential equations in  \eqref{eq:ZeDifferentialEquation} can be computed explicitly.

\section*{Acknowledgments}

YT thanks LPTMS for hospitality.
This work has benefitted from the financial support \og Investissements d'Avenir du LabEx PALM \fg{} (ANR-10-LABX-0039-PALM), project ProMAFluM.
  

\appendix

\section{Relation between \eqref{eq:BoundaryConditions} and \eqref{eq:limitPhiHat}}
\label{app:FourierReciprocity}
We start from \eqref{eq:SplitPhiHat}, 
where $\phi(z;\Lambda)$ is bounded and decays as $\phi(z;\Lambda)\simeq A\,|z|^{-2-q}$ asymptotically.
The first integral in \eqref{eq:SplitPhiHat} is a constant.
We now introduce a large positive number, say $z_c$, and write the second integral as a sum of two integrals: one over the interval $-z_c < z <z_c$ and the other over its complement.
The first of these integrals clearly exhibits an analytic behaviour
$\int_{|z|<z_c}\D z\,\phi(z;\Lambda) \left(1-\EXP{-\I sz}\right)=\alpha_1\,s  + \alpha_2\,s^2+\cdots$
as $s\to0$. As for the second of the integrals, we assume that $z_c$ is so large that the power tail dominates in the integrand:
\begin{align}
&A\int_{|z|>z_c}\D z\,\left(1-\EXP{-\I sz}\right)\,|z|^{-2-q}
\nonumber\\
=& A\,|s|^{q+1} 
\int_{|y|>z_c|s|}\D y\,\left(1-\EXP{-\I y}\right)\,|y|^{-2-q}\,.
 \label{eq:AppendixA1}
\end{align}
This last integral is convergent for $q>-1$.
Obviously, when $q<1$ the integral over $y$ has a limit as $s\to0$. 
For finite $s$, the expansion of the integral in the right hand side of \eqref{eq:AppendixA1} produces the non analytic series $\beta_1\,|s|^{q+1}  + \beta_2\,|s|^{q+2}+\cdots$.
This establishes the correspondence between the power law tail of $\phi(z;\Lambda)$ as $z\to\pm\infty$ and the non analytic $|s|^{q+1}$ term in its Fourier transform.
Such a relation, which characterizes Fourier reciprocity, is known as a ``Tauberian theorem'' in the mathematical literature.



\begin{thebibliography}{10}
\bibitem{footnote1}
The Dyson paper is devoted to the study of the random spring chain. Nevertheless, the model (denoted ``type I") solved in the paper involves artificial correlations between the spring constants and/or masses. This model coincides with the tight-binding model with random couplings. See \cite{For21} for a wide-ranging discussion of this paper.
\bibitem{footnote3}
The recursive approach was revisited and generalized by Monthus in Ref.~\cite{Mon21}, and the relevance of recent progresses in the theory of large deviations of random processes was pointed out.
\bibitem{footnote5}
The limit law exists if $\mean{\ln^+|v_n|}<\infty$ \cite{CriPalVul93}, where $\ln^+(x)=\ln x$ for $x>1$ and $\ln^+(x)=0$ otherwise.
\bibitem{footnote6} 
The case where $\levy(s)/s\sim s^{\alpha-1}$ with $\alpha\in]0,2[$, i.e. may diverge, has been studied in \cite{Tex20b}.
\bibitem{footnote6bis}
The utility of various integral transforms in the practical calculation of the Lyapunov exponent was discussed in considerable detail in Refs.~\cite{ComTexTou10,ComTexTou11,ComLucTexTou13,GraTexTou14,Tex20,ComTexTou19}.  The reason for the effectiveness of the Fourier transform for our particular model is that delta interactions are associated with the subgroup of $\text{SL}(2,{\mathbb R})$ consisting of upper triangular matrices, see Eq.~\eqref{eq:MatricesOfInterest}; in Fourier space, the infinitesimal generator of this subgroup becomes a simple multiplication operator.
\bibitem{footnote7}
The form \eqref{eq:GLEUniversalLimit} was generalized in \cite{Tex20b} to the case of power-law disorder with $p(v)\sim|v|^{-1-\alpha}$ with $\alpha\in]0,2[$ (Cauchy is $\alpha=1$). In this case the GLE is defined for $q\in]-2-\alpha,\alpha[$.
\bibitem{footnote8}
  We recall that $\psi(x)$ is the solution of the initial-value (Cauchy) problem and not a normalised eigenfunction.
\end{thebibliography}

\end{document}